\documentclass[aps,prl,twocolumn,superscriptaddress]{revtex4-1}

\usepackage{graphicx}
\usepackage{dcolumn}
\usepackage{bm}
\usepackage{float}
\usepackage{lipsum}
\usepackage{amsmath}
\usepackage{booktabs}

\begin{document}


\title{Collisionless shock acceleration of narrow energy spread ion beams from mixed species plasmas using 1 $\mu$m lasers
}


\author{A. Pak}
\email{pak5@llnl.gov}
\affiliation{Lawrence Livermore National Laboratory, Livermore, CA, 94550, USA}
\author{S. Kerr}
\affiliation{Department of Electrical and Computer Engineering, University of Alberta, Edmonton, Alberta T6G 2V4, Canada}

\author{N. Lemos}
\author{A. Link}
\author{P. Patel}
\author{F. Albert}
\author{L. Divol}
\author{B. B. Pollock}
\affiliation{Lawrence Livermore National Laboratory, Livermore, CA, 94550, USA}

\author{D. Haberberger}
\author{D. Froula}
\affiliation{Laboratory for Laser Energetics, University of Rochester, 250 East River Road, Rochester, New York 14623-1299, USA}

\author{M. Gauthier}
\author{S. H. Glenzer}
\affiliation{SLAC National Accelerator Laboratory, Menlo Park, CA 94025, USA}

\author{A. Longman}
\author{L. Manzoor}
\author{R. Fedosejevs}
\affiliation{Department of Electrical and Computer Engineering, University of Alberta, Edmonton, Alberta T6G 2V4, Canada}

\author{S. Tochitsky}
\author{C. Joshi}
\affiliation{Department of Electrical Engineering, UCLA, Los Angeles, California 90095, USA}

\author{F. Fiuza}
\email{fiuza@slac.stanford.edu}
\affiliation{SLAC National Accelerator Laboratory, Menlo Park, CA 94025, USA}



\date{\today}
\begin{abstract}
Collisionless shock acceleration of protons and C$^{6+}$
ions has been achieved by the interaction of a 10$^{20}$ W/cm$^2$, 1 $\mu$m laser with a near-critical density plasma. Ablation of the initially solid density target by a secondary laser allowed for systematic control of the plasma profile. This enabled the production of beams with peaked spectra with energies of 10-18 MeV/a.m.u. and energy spreads of 10-20$\%$ with up to 3x10$^9$ particles within these narrow spectral features. The narrow energy spread and similar velocity of ion species with different charge-to-mass ratio are consistent with acceleration by the moving potential of a shock wave. Particle-in-cell simulations show shock accelerated beams of protons and C$^{6+}$ ions with energy distributions consistent with the experiments. Simulations further indicate the plasma profile determines the trade-off between the beam charge and energy and that with additional target optimization narrow energy spread beams exceeding 100 MeV/a.m.u. can be produced using the same laser conditions. 
\end{abstract}

\pacs{}

\maketitle

\section{I. Introduction}
The ability to study the properties of high energy density matter in the laboratory is expanding our understanding of the physics associated with inertial fusion targets, planetary interiors, and astrophysical systems \cite{Kraus2017,Hurricane2016,Ross2017}. Laser-produced ion beams have proven an invaluable tool for both creating and probing such high energy density matter \cite{clark2000,Patel2003,Dyer2008,Fernandez2014,Nilson2006}.  Traditionally, these beams have been accelerated via the target normal sheath acceleration (TNSA) mechanism which produces a continuous exponentially decreasing energy spectrum \cite{Snavely2000}. In the pursuit of new applications and increased precision, significant effort has gone into exploring other schemes to extend the maximum ion energy and reduce the energy spread to ~1-10$\%$ \cite{Tikhonchuk2005,Bulanov2008,Esirkepov2004,Hegelich2005,Palmer2011,Henig2009,Kar2012,Palaniyappan2015}. Recently, proof-of-principle experiments have shown that such narrow energy spread proton beams, containing $2\times10^{5}$ particles, can be accelerated up to $\sim$20 MeV in tailored near-critical density plasmas via an electrostatic shock wave driven in a hydrogen gas jet plasma by a 10 $\mu$m CO$_2$ laser \cite{Haberberger2012,Tresca2015}. While these results are promising, CO$_2$ lasers are not commonly available. Furthermore, it is desirable to produce beams with higher charge and particle energy, which generally requires operating at higher densities and intensities.  This can only be achieved by using more ubiquitous solid-state high intensity lasers at a wavelength of $\sim$1 $\mu$m. 

Here, we report for the first time on collisionless shock acceleration (CSA) experiments with a 1 $\mu$m laser that produced proton and ion beams with narrow energy spreads $\Delta$E/E of 10-20$\%$ centered at 10-18 MeV/a.m.u. and with a total number of particles in these peaks up to $3\times10^9$. To produce a plasma density profile suitable for CSA, we have used a secondary laser to ablate a Mylar (C$_{10}$H$_8$O$_4$) foil. For this profile we observed similar velocity distributions of accelerated protons and heavier ions, consistent with the reflection from the moving potential of an electrostatic shock. The number of particles within the narrow distributions of accelerated ions is $\sim 10^4 \times$ larger than obtained in previous CSA experiments\cite{Haberberger2012}. Two-dimensional (2D) particle-in-cell (PIC) simulations that model the laser interaction with a CH plasma for the experimental conditions show CSA of multiple ion species with spectra consistent with observations. Analysis of simulation results reveal that the plasma density profile determines the trade-off between energy gain and number of accelerated particles, by controlling the velocities of the shock and of the expanding plasma. This suggests that further control over the density profile could allow beams to be tuned according to application needs.  
\begin{figure}
\includegraphics{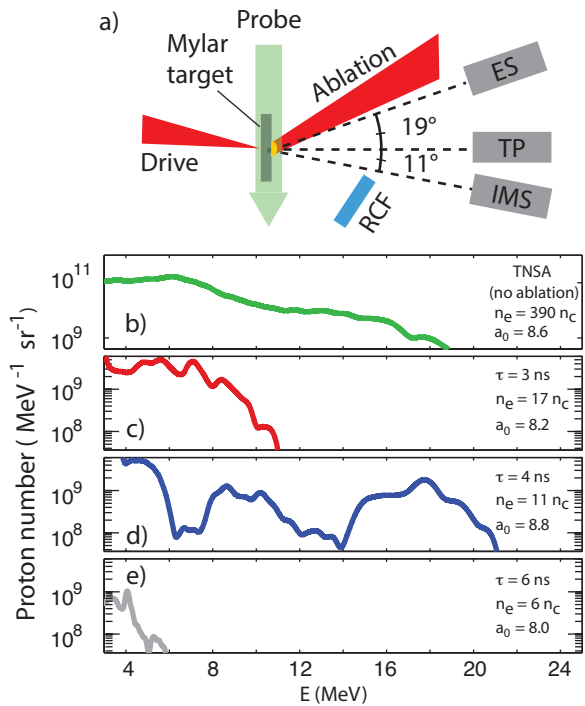}%
\caption{\label{setup} (Color online) a)Experimental setup.  A near-critical density target is created by first irradiating a Mylar foil with an ablation laser. After the target has expanded for a time $\tau$, a high-intensity ps duration laser pulse is focused onto the target to produce the electrostatic shock wave.  Accelerated ions are measured by the imaging magnetic spectrometer (IMS) and Thomson parabola (TP). TP measurements indicate the majority of accelerated ions are C$^{6+}$/O$^{8+}$.  Accelerated electrons are measured by a permanent magnet electron spectrometer.  Radiochromic film (RCF) was used to measure a portion of the spatial beam profile.  Orthogonal to the target, a probe laser was used to measure the target expansion. Accelerated proton spectra shown for b) an unablated foil, and c-e) at different time delays from consecutive shots. The inferred peak $n_e$ of the target and laser $a_0$ are also denoted. Only signal $>4\times$ the background variation is shown.  }
\end{figure}
\section{II. Collisionless shock formation }
The formation of a collisionless electrostatic shock requires the creation of a localized region of higher pressure within a plasma with $T_e \gg T_i$ \cite{Silva2004,Fiuza2012,Medvedev2014}. The interaction of a high-intensity laser with a near-critical density plasma can efficiently produce these conditions\cite{Fiuza2012}. 
As this region of high pressure expands, it can drive a shock wave with velocity $v_s=M_s C_s$ into the surrounding lower pressure plasma. Here $v_s$ is defined in the upstream plasma frame, $M_s$ denotes the shock Mach number, and for $T_e \gg T_{\textrm{i}}$ the ion sound speed $C_s = (Z T_e / m_i)^{1/2}$ depends on the electron temperature, $T_e$, the ion mass, $m_i$, and the ion charge state, $Z$, of the plasma. The shock can reflect upstream ions if its electrostatic potential, $Ze\Phi$, is larger than the kinetic energy of the in-flowing ions, i.e. $\overline{\Phi} = Z e \Phi / (\frac{1}{2}m_i v_{s}^2) > 1$. Provided this criterion is satisfied, the shock can reflect ions of different charge-to-mass ($Z/m_p A$) ratios to a velocity $2 v_{s}$. The final ion velocity will result from contributions from both shock reflection and sheath acceleration (which depends on the plasma profile) and can be written as $v_{f}=2v_{s}+v_{\rm{sheath}}$.

To produce high-energy ($\gtrsim$ 10 MeV/a.m.u.) ion beams with moderate strength shocks ($1<M_s<3)$, the plasma needs to be heated to $T_e \gtrsim$ 1 MeV to drive a shock with $v_s \gtrsim$ 0.1 c. Balancing the energy density of the laser with that of the target, $T_e$ can be estimated as \cite{Fiuza2012},
\begin{equation}
T_e [\mathrm{MeV}]=2.6\eta a_0^2 \frac{n_{\textrm{c}}}{n_e} \frac{\tau_0 [\mathrm{ps}]}{L [10 \mathrm{\mu m}]}
\label{ebal}
\end{equation} 
where $\tau_0$ is the laser pulse duration, $L$ is the target thickness, $n_e$ is the electron density, $n_c \approx 10^{21} cm^{-3}$ is the critical density for 1 $\mu$m light and $a_0 = \frac{e A}{m_e c^2}$ is the normalized vector potential of the laser.  At high laser intensity ($a_0 \gg 1$) the coupling of the laser to the target, $\eta$, can be optimized to values of $\sim 0.5$ for a peak electron density near the relativistic critical density $n_e = a_0 n_c$\cite{Gong2016}.  To explore CSA in this high intensity regime, experiments were performed at the Titan laser facility.
\section{III. Experiment}
As seen in Fig. \ref{setup} a), to produce a near-critical density target, the 0.5 $\mu$m thick Mylar foil was first irradiated by the 10 ns long, 1 $\mu$m wavelength ablation laser focused to a diameter of $\sim$550 $\mu$m and an average peak intensity 1.2$\times$10$^{11}$ W/cm$^2$.  This approach was pursued in order to produce plasmas with peak densities of $\sim$10$n_c$ and lengths $L\leq50 \mu$m required to obtain T$_e>$ 1 MeV using a drive laser $a_0\sim10$ per Eq. \eqref{ebal}. The ablation of material creates a density gradient and an associated quasi-uniform sheath field that allows the shock reflected ions to exit the target with their narrow energy spread largely preserved\cite{Grismayer2008,Fiuza2013}. After the target expansion, an high-intensity drive laser, with a wavelength of 1 $\mu$m and a duration of $\sim$1 ps, was focused onto the plasma to generate the shock wave. The longitudinal position of the target was varied by up to 150 $\mu$m. This changed the full width half maximum (FWHM) of the laser spot from 5-9 $\mu$m and peak $a_0$ from $\sim 4.5-8.5$, respectively.

\begin{figure}
\includegraphics{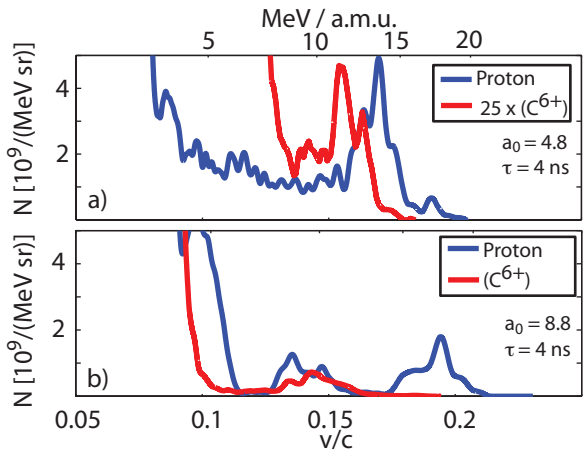}%
\caption{(Color online) a-b) The observed proton and ion velocity spectrum measured at the optimal delay of $\tau$ = 4 ns.  In a) and b) the drive laser power was held constant while the laser focus was varied, changing the incident $a_0$ from 4.8 to 8.8 , respectively.}
\label{ProtonEionV}
\end{figure}

To optimize the CSA process, the peak plasma density and profile were changed by varying the delay, $\tau$, between the beginning of the ablation laser and the high-intensity short pulse drive laser. Shadowgraphic measurements of the foil ablation\cite{Supp}, were found to be consistent with radiation hydrodynamic calculations using the code HYDRA\cite{HYDRA}.  These calculations indicate that the peak density decreases from 16.7 to 6.1 $n_c$ and the FWHM target thickness increases from 18 to 44 $\mu$m as $\tau$ was increased from 3 to 6 ns.   
An imaging magnetic spectrometer (IMS)\cite{Chen2010} was used to measure the accelerated ion spectrum along the axis of laser propagation.  The measured proton spectra as a function of delay between the ablation and drive laser are shown in Fig. \ref{setup} b-e).  With the ablation laser off, the proton spectrum is characteristic of the TSNA mechanism and extends to maximum of $\sim$ 19 MeV (Fig. \ref{setup} b)). With the ablation laser on, the delay was then increased on consecutive shots.  For $\tau=3$ ns, Fig. \ref{setup} c), the cutoff energy decreases to 11 MeV.  This is attributed to the increasing rear scale length of the target, which reduces the TNSA field.  Interestingly, at $\tau=4$ ns, Fig. \ref{setup} d) shows that two spectrally narrow and distinct peaks appear at $\sim$10 and $\sim$18 MeV, respectively, in contrast to the usual TNSA continuum.  This suggests an additional acceleration mechanism is present and capable of accelerating narrow distributions of protons to energies comparable to the maximum TNSA cutoff energy. Compared to the unablated foil, at $\tau = 4$ ns, the number of escaping electrons was observed to increase 4$\times$\cite{Supp}.This is consistent with the increased laser coupling and heating required to produce CSA protons. For $\tau=6$ ns, Fig. \ref{ProtonEionV} e), no protons $>$5 MeV were observed.   These results clearly show an optimal acceleration regime at $\tau = 4$ ns, indicating that the production of narrow energy distributions is sensitive to the plasma density profile at this time.    
\begin{figure*}
\includegraphics[width=1.0\textwidth]{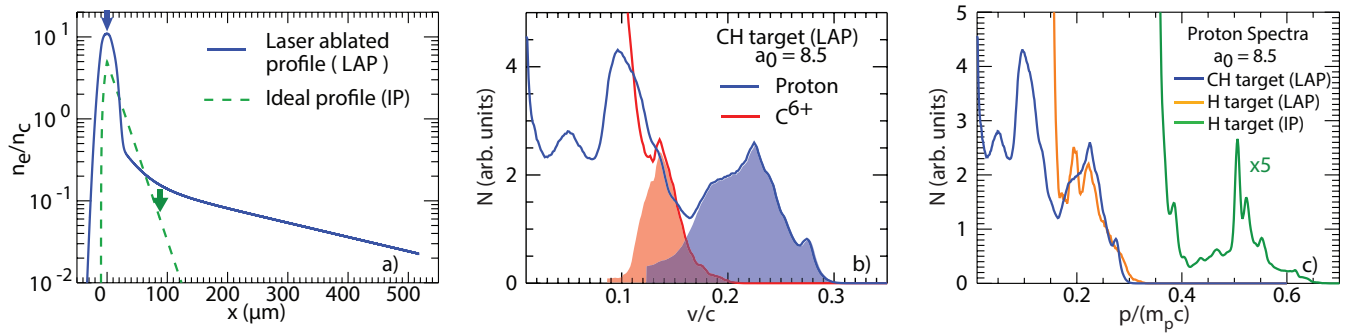}%
\caption{\label{sim} (Color online) Ion spectra produced from PIC simulations using different plasma profiles and ion compositions. a)Initial electron density profile, with the dashed and solid curves denoting the theoretical idealized profile (IP)\cite{Fiuza2012} and the expected laser ablated profile (LAP) of the experiment at $\tau$=4 ns from HYDRA calculations, respectively.  In these simulations a 1 ps laser with $a_0=8.5$ irradiates the target from the left to produce the shock wave. For each profile the location at which shock reflection begins is denoted by an arrow. b) The proton and C$^{6+}$ spectrum produced from a CH plasma with a LAP.  The shaded region denotes the shock reflected portion of the spectra as identified from the ion velocity phase space. c) Comparison of the proton spectra obtained with the same laser for a CH target with a LAP (blue), a pure H target with a LAP (orange), and a pure H target with a IP (green, with amplitude multiplied by 5). }
\end{figure*}

Figure 2 and additional spectra detailed in the supplemental material show that at $\tau$ = 4 ns, spectra with narrow peaks of protons were consistently observed at energies between 7.9 and 17.7 MeV.  At this delay, a narrow distribution of heavier ions with a peak velocity within 30$\%$ of the proton peak velocity was also consistently seen.  The observation of multiple species of ions with different charge-to-mass ratios being accelerated to similar velocities and into narrow distributions is consistent with the reflection and acceleration from a moving potential associated with a collisionless shock and not expected to result from TNSA. Differential filtering of the IMS image plate detector allowed for discrimination between proton and heavier ion spectral features\cite{Supp}. Due to having the same Z/A ratio, differentiating between C$^{6+}$ and O$^{8+}$ ions using the IMS or TP was not possible. The ion signal is assumed to be comprised predominantly of C$^{6+}$ ions as the Mylar target has 2.5X more carbon than oxygen ions.

While the production of narrow distributions of protons and ions at similar velocities was consistently observed at a $\tau$ = 4 ns, the spectral shape and peak energy of these distributions was observed to vary shot to shot and to be sensitive to the incident laser spot size.  Shot to shot the energy and energy spread of the higher velocity peak was observed to vary between 11.3-17.7 MeV and 8.5-15.8$\%$, respectively, as the incident a$_0$ was varied between 8.1 and 8.8.  Additionally, as seen in Fig 2 a) and b), at the same incident laser power the spectral distribution of protons was observed to change when the spot size was increased and the incident a$_0$ reduced to 4.8.  The variation in energy and spectral shape is thought to arise from differences in laser-plasma coupling and heating. This is influenced by shot to shot variations in the laser focusing and resulting intensity due to the thermal lensing of the laser, the plasma density profile, self-focusing and target alignment.   Similar energy variation is common in other high-intensity laser plasma acceleration schemes \cite{Fuchs2005}.  

Within the FWHM of the proton peaks at 13.5 and 17.7 MeV observed in Fig. \ref{ProtonEionV} a-b), the total number of protons was estimated to be 3.2$\pm$0.9$\times10^{9}$ and 1.0$\pm0.6\times10^{9}$, respectively.  Measurements at these conditions show a proton beam divergence of $\sim$24$^\circ$.    The number of accelerated protons observed is substantially higher ($\sim 10^4\times$) than obtained in previous CSA experiments conducted at lower densities and intensities with 10 $\mu$m wavelength lasers. Moreover, the higher energy, narrow spread, peak shown in Fig. \ref{ProtonEionV} b) contains similar ($\sim$80$\%$) charge to the TNSA beam shown in Fig. \ref{setup} b) at the same energy and bandwidth. This shows that CSA can represent a significant advantage for applications requiring narrow energy spread beams, since it would avoid the beam transmission losses and added complexity associated with energy selection techniques of broadband TNSA beams. Experiments using magnetic-field based techniques to reduce the bandwidth of TNSA beams have been limited to 0.1$\%$ transmission efficiencies\cite{Toncian2006,Schollmeier2008}. Recent simulations of more advanced electro-optics indicate that under optimal conditions the transmission efficiency at 60-200 MeV can approach 5-20$\%$ \cite{SchillaciJINST,Masood2014}. At energies between 5 and 8 MeV, the transmission through a set of four quadrapoles was inferred to range between $\sim$15-100$\%$ \cite{SchillaciJINST2}.
\section{IV. Simulations}
In order to better understand how the laser-ablated density profile (LAP) and multi-ion species plasma impact the scaling of ion acceleration with laser intensity, 2D PIC simulations with OSIRIS 3.0 \cite{Fonseca2002} were performed. The simulations modeled the interaction of the drive laser pulse ($\lambda_0 = 1 \mu$m, $\tau_0 = 1$ ps, and $a_0 = 8.5$) with a CH plasma for the experimentally expected profile obtained with HYDRA at $\tau=4$ ns (Fig. \ref{sim} a)). The LAP has a peak density $n_e = 11 n_c$ and a FWHM thickness $L = 25 \mu$m, followed by a long low-density exponential profile with a scale-length of $L_g = 250 \mu$m at the rear side. In order to simulate the temporal dynamics of the interaction, a long and narrow simulation box was used that extended 830 $\mu$m and 10 $\mu$m in the direction along and transverse to the laser propagation, respectively\cite{sim}.  
  
The simulations confirm the formation of an electrostatic shock with $v_s \sim 0.045 c$, that reflects both protons and $C^{6+}$ ions from the upstream plasma to $\sim 0.09 c$. Shock reflection starts near the peak density of the plasma (blue arrow in Fig. \ref{sim} a)) soon after the laser reaches peak intensity. The sharp change in the density profile near $n_e \sim 0.5 n_c$ at the rear side of the plasma (where the ablation laser is absorbed) leads to the generation of a localized space-charge electric field. As shock reflected protons ($C^{6+}$ ions) experience this field they gain an additional velocity $v_{\rm{sheath}} \sim 0.13 c  ~(\sim0.05 c)$. The differences in $v_{\rm{sheath}}$ are mostly due to the different Z/A ratio of the two species. After this region, the typical TNSA field is strongly suppressed due to the long density scale-length, and the maximum velocity remains the same. This leads to a final velocity of the $C^{6+}$ ions within $\sim 35\%$ of the proton velocity, similar to the experiments. Moreover, the final particle spectra obtained is also consistent with the experimental observations, showing peaks with energies (and energy spreads) of $23$ MeV ($\Delta E/E = 64\%$) for protons and $9$ MeV/a.m.u. ($\Delta E/E = 33\%$) for $C^{6+}$ ions (Fig. \ref{sim} b)). The energy spread is mostly determined by the temporal evolution of the shock, which slows down due to dissipation by ion reflection \cite{Macchi2012}. The (slice) energy spread at each reflection point is significantly smaller, 19$\%$ for C$^{6+}$ and 8$\%$ for protons. 

Simulations conducted with the same laser and electron density profile, but with a pure hydrogen plasma, show that the spectrum of reflected protons is very similar to the case of a CH plasma (Fig. \ref{sim} c)). This indicates that the presence of multiple ion species does not significantly affect the maximum obtainable velocity. However, the presence of multiple ion species is found to change the expansion dynamics downstream of the shock, inducing modulations in the lower energy portion of the spectrum, as seen in Fig. \ref{sim} b) and in some of the experimental results. This will be discussed in more detail elsewhere. 

The impact of the experimental LAP on particle acceleration was investigated by comparing these results with those obtained in simulations where the same laser interacts with a hydrogen target with the theoretical ideal profile (IP) discussed in Ref. \cite{Fiuza2012}. The IP has a sharp linear rise over $10\mu$m on the front side, followed by a exponential profile on the rear side with a scale-length $L_g = 20 \mu$m (Fig. \ref{sim} a)). The FWHM thickness of the target is $L = 17.5 \mu$m. For a fixed density profile, it was found that a peak density $n_e = 5 n_c$ maximizes the energy gain by CSA. For these conditions, the laser absorption and electron temperature is higher than with the LAP, as described by Eq. \eqref{ebal}. An electrostatic shock is formed with $v_s \sim 0.145 c$. At such high velocity, the shock cannot efficiently reflect the upstream protons initially. In this case, CSA requires the upstream protons to be first accelerated in the controlled TNSA field, which reduces their kinetic energy in the shock frame. For the density scale-length $L_g = 20 \mu$m, protons acquire $v_{\rm{sheath}} \sim 0.22  c$, before they are reflected by the shock. The final proton beam energy is $E = 113$ MeV with $\Delta E/E = 4\%$ (Fig. \ref{sim} c)), consistent with the CSA energy scaling \cite{Fiuza2012}. While the energy obtained with the IP is significantly higher, the total number of protons contained in the reflected beam is $\sim 30\times$ smaller than in the LAP. This is because efficient reflection only begins at the rear side of the target near $n_e \sim 0.1 n_c$ as seen in Fig. \ref{sim} a).

These results indicate that the plasma profile controls both the charge and energy of CSA beams. Laser ablation of thinner foils ($<$0.5 $\mu$m) may allow the production of plasmas with $n_e\sim$5 $n_c$ and L $\sim$17.5 $\mu$m that, with the laser used in these experiments, is estimated to produce $\sim$80 MeV proton beams. Simulations indicate that further tuning of the rear-side density scale-length to $L_g\sim$20 $\mu$m would produce proton beams with $>$100 MeV, but with less charge. 
\section{V. Conclusions}
In conclusion, we report on the first experimental evidence of efficient CSA of narrow distributions of protons and heavier ions using a high-intensity 1 $\mu$m wavelength laser with a peak $a_0$ $\sim$8.5. By tuning the plasma profile using laser-ablation, beams with energies up to ~18 MeV/a.m.u., energy spreads of $10-20\%$, containing up to $3\times10^9$ particles were produced. The number of particles in these distributions was 10$^4 \times$ higher than previous CSA work conducted with 10 $\mu$m wavelength laser systems. These results demonstrate the ability of CSA to efficiently accelerate high yield, narrow distributions of ions to meet the needs of applications.  Additionally, the simultaneous acceleration of ion beams with different Z/A ratios to similar velocities offers a promising source for more accurately diagnosing the electromagnetic fields of high-energy-density plasmas. Results from PIC simulations are consistent with the experimental data and reveal that the control of the plasma profile allows the optimization of the beam charge or energy, depending on the application needs. Precise shaping of near-critical density plasma profiles would allow the generation of $> 100$ MeV/a.m.u. with the same laser system. This could be achieved in the future by reducing the foil thickness, by changing the wavelength of the photons used to ablate the target (e.g. x-rays), or by directly fabricating the profile via 3D printing.
 
We wish to thank H. Chen, and A. Hazi for their support of these experiments, and C. Curry, A. Kemp and C. Roedel for valuable discussions.  This work and the use of the Jupiter Laser Facility was performed under the auspices of the U.S. Department of Energy (DOE) under Contract No. DE-AC52-07NA27344, with support from the LDRD Program (15-LW-095).  
Additional support was provided by U.S. DOE under contract numbers DE- AC02-76SF00515,DE-SC0010064,and DE-NA0001944,  the DOE Office of Science, FES under FWP 100237, FWP 100182, FWP 100331 and SCW1575-1, NNSA grant no. DE-NA0002950, NSF grant 1734315, and NSERC grant no. RGPIN-2014-05736. The work at UCLA was supported by NNSA grant  DE-NA0003873. The authors acknowledge the OSIRIS Consortium, consisting of UCLA and IST (Portugal) for the use of the OSIRIS 3.0 framework and the visXD framework. Simulations were conducted on Mira (ALCF) through an ALCC award and on Vulcan (LLNL). 

\bibliography{references2}

\begin{thebibliography}{37}%
\makeatletter
\providecommand \@ifxundefined [1]{%
 \@ifx{#1\undefined}
}%
\providecommand \@ifnum [1]{%
 \ifnum #1\expandafter \@firstoftwo
 \else \expandafter \@secondoftwo
 \fi
}%
\providecommand \@ifx [1]{%
 \ifx #1\expandafter \@firstoftwo
 \else \expandafter \@secondoftwo
 \fi
}%
\providecommand \natexlab [1]{#1}%
\providecommand \enquote  [1]{``#1''}%
\providecommand \bibnamefont  [1]{#1}%
\providecommand \bibfnamefont [1]{#1}%
\providecommand \citenamefont [1]{#1}%
\providecommand \href@noop [0]{\@secondoftwo}%
\providecommand \href [0]{\begingroup \@sanitize@url \@href}%
\providecommand \@href[1]{\@@startlink{#1}\@@href}%
\providecommand \@@href[1]{\endgroup#1\@@endlink}%
\providecommand \@sanitize@url [0]{\catcode `\\12\catcode `\$12\catcode
  `\&12\catcode `\#12\catcode `\^12\catcode `\_12\catcode `\%12\relax}%
\providecommand \@@startlink[1]{}%
\providecommand \@@endlink[0]{}%
\providecommand \url  [0]{\begingroup\@sanitize@url \@url }%
\providecommand \@url [1]{\endgroup\@href {#1}{\urlprefix }}%
\providecommand \urlprefix  [0]{URL }%
\providecommand \Eprint [0]{\href }%
\providecommand \doibase [0]{http://dx.doi.org/}%
\providecommand \selectlanguage [0]{\@gobble}%
\providecommand \bibinfo  [0]{\@secondoftwo}%
\providecommand \bibfield  [0]{\@secondoftwo}%
\providecommand \translation [1]{[#1]}%
\providecommand \BibitemOpen [0]{}%
\providecommand \bibitemStop [0]{}%
\providecommand \bibitemNoStop [0]{.\EOS\space}%
\providecommand \EOS [0]{\spacefactor3000\relax}%
\providecommand \BibitemShut  [1]{\csname bibitem#1\endcsname}%
\let\auto@bib@innerbib\@empty
\bibitem [{\citenamefont {Kraus}\ \emph {et~al.}(2017)\citenamefont {Kraus},
  \citenamefont {Vorberger}, \citenamefont {Pak}, \citenamefont {Hartley},
  \citenamefont {Fletcher}, \citenamefont {Frydrych}, \citenamefont {Galtier},
  \citenamefont {Gamboa}, \citenamefont {Gericke}, \citenamefont {Glenzer},
  \citenamefont {Granados}, \citenamefont {MacDonald}, \citenamefont
  {MacKinnon}, \citenamefont {McBride}, \citenamefont {Nam}, \citenamefont
  {Neumayer}, \citenamefont {Roth}, \citenamefont {Saunders}, \citenamefont
  {Schuster}, \citenamefont {Sun}, \citenamefont {van Driel}, \citenamefont
  {D{\"o}ppner},\ and\ \citenamefont {Falcone}}]{Kraus2017}%
  \BibitemOpen
  \bibfield  {author} {\bibinfo {author} {\bibfnamefont {D.}~\bibnamefont
  {Kraus}}, \bibinfo {author} {\bibfnamefont {J.}~\bibnamefont {Vorberger}},
  \bibinfo {author} {\bibfnamefont {A.}~\bibnamefont {Pak}}, \bibinfo {author}
  {\bibfnamefont {N.~J.}\ \bibnamefont {Hartley}}, \bibinfo {author}
  {\bibfnamefont {L.~B.}\ \bibnamefont {Fletcher}}, \bibinfo {author}
  {\bibfnamefont {S.}~\bibnamefont {Frydrych}}, \bibinfo {author}
  {\bibfnamefont {E.}~\bibnamefont {Galtier}}, \bibinfo {author} {\bibfnamefont
  {E.~J.}\ \bibnamefont {Gamboa}}, \bibinfo {author} {\bibfnamefont {D.~O.}\
  \bibnamefont {Gericke}}, \bibinfo {author} {\bibfnamefont {S.~H.}\
  \bibnamefont {Glenzer}}, \bibinfo {author} {\bibfnamefont {E.}~\bibnamefont
  {Granados}}, \bibinfo {author} {\bibfnamefont {M.~J.}\ \bibnamefont
  {MacDonald}}, \bibinfo {author} {\bibfnamefont {A.~J.}\ \bibnamefont
  {MacKinnon}}, \bibinfo {author} {\bibfnamefont {E.~E.}\ \bibnamefont
  {McBride}}, \bibinfo {author} {\bibfnamefont {I.}~\bibnamefont {Nam}},
  \bibinfo {author} {\bibfnamefont {P.}~\bibnamefont {Neumayer}}, \bibinfo
  {author} {\bibfnamefont {M.}~\bibnamefont {Roth}}, \bibinfo {author}
  {\bibfnamefont {A.~M.}\ \bibnamefont {Saunders}}, \bibinfo {author}
  {\bibfnamefont {A.~K.}\ \bibnamefont {Schuster}}, \bibinfo {author}
  {\bibfnamefont {P.}~\bibnamefont {Sun}}, \bibinfo {author} {\bibfnamefont
  {T.}~\bibnamefont {van Driel}}, \bibinfo {author} {\bibfnamefont
  {T.}~\bibnamefont {D{\"o}ppner}}, \ and\ \bibinfo {author} {\bibfnamefont
  {R.~W.}\ \bibnamefont {Falcone}},\ }\href {\doibase
  10.1038/s41550-017-0219-9} {\bibfield  {journal} {\bibinfo  {journal} {Nature
  Astronomy}\ }\textbf {\bibinfo {volume} {1}},\ \bibinfo {pages} {606}
  (\bibinfo {year} {2017})}\BibitemShut {NoStop}%
\bibitem [{\citenamefont {Hurricane}\ \emph {et~al.}(2016)\citenamefont
  {Hurricane}, \citenamefont {Callahan}, \citenamefont {Casey}, \citenamefont
  {Dewald}, \citenamefont {Dittrich}, \citenamefont {Döppner}, \citenamefont
  {Haan}, \citenamefont {Hinkel}, \citenamefont {Berzak~Hopkins}, \citenamefont
  {Jones}, \citenamefont {Kritcher}, \citenamefont {Le~Pape}, \citenamefont
  {Ma}, \citenamefont {MacPhee}, \citenamefont {Milovich}, \citenamefont
  {Moody}, \citenamefont {Pak}, \citenamefont {Park}, \citenamefont {Patel},
  \citenamefont {Ralph}, \citenamefont {Robey}, \citenamefont {Ross},
  \citenamefont {Salmonson}, \citenamefont {Spears}, \citenamefont {Springer},
  \citenamefont {Tommasini}, \citenamefont {Albert}, \citenamefont {Benedetti},
  \citenamefont {Bionta}, \citenamefont {Bond}, \citenamefont {Bradley},
  \citenamefont {Caggiano}, \citenamefont {Celliers}, \citenamefont {Cerjan},
  \citenamefont {Church}, \citenamefont {Dylla-Spears}, \citenamefont {Edgell},
  \citenamefont {Edwards}, \citenamefont {Fittinghoff}, \citenamefont
  {Barrios~Garcia}, \citenamefont {Hamza}, \citenamefont {Hatarik},
  \citenamefont {Herrmann}, \citenamefont {Hohenberger}, \citenamefont
  {Hoover}, \citenamefont {Kline}, \citenamefont {Kyrala}, \citenamefont
  {Kozioziemski}, \citenamefont {Grim}, \citenamefont {Field}, \citenamefont
  {Frenje}, \citenamefont {Izumi}, \citenamefont {Gatu~Johnson}, \citenamefont
  {Khan}, \citenamefont {Knauer}, \citenamefont {Kohut}, \citenamefont
  {Landen}, \citenamefont {Merrill}, \citenamefont {Michel}, \citenamefont
  {Moore}, \citenamefont {Nagel}, \citenamefont {Nikroo}, \citenamefont
  {Parham}, \citenamefont {Rygg}, \citenamefont {Sayre}, \citenamefont
  {Schneider}, \citenamefont {Shaughnessy}, \citenamefont {Strozzi},
  \citenamefont {Town}, \citenamefont {Turnbull}, \citenamefont {Volegov},
  \citenamefont {Wan}, \citenamefont {Widmann}, \citenamefont {Wilde},\ and\
  \citenamefont {Yeamans}}]{Hurricane2016}%
  \BibitemOpen
  \bibfield  {author} {\bibinfo {author} {\bibfnamefont {O.~A.}\ \bibnamefont
  {Hurricane}}, \bibinfo {author} {\bibfnamefont {D.~A.}\ \bibnamefont
  {Callahan}}, \bibinfo {author} {\bibfnamefont {D.~T.}\ \bibnamefont {Casey}},
  \bibinfo {author} {\bibfnamefont {E.~L.}\ \bibnamefont {Dewald}}, \bibinfo
  {author} {\bibfnamefont {T.~R.}\ \bibnamefont {Dittrich}}, \bibinfo {author}
  {\bibfnamefont {T.}~\bibnamefont {Döppner}}, \bibinfo {author}
  {\bibfnamefont {S.}~\bibnamefont {Haan}}, \bibinfo {author} {\bibfnamefont
  {D.~E.}\ \bibnamefont {Hinkel}}, \bibinfo {author} {\bibfnamefont {L.~F.}\
  \bibnamefont {Berzak~Hopkins}}, \bibinfo {author} {\bibfnamefont
  {O.}~\bibnamefont {Jones}}, \bibinfo {author} {\bibfnamefont {A.~L.}\
  \bibnamefont {Kritcher}}, \bibinfo {author} {\bibfnamefont {S.}~\bibnamefont
  {Le~Pape}}, \bibinfo {author} {\bibfnamefont {T.}~\bibnamefont {Ma}},
  \bibinfo {author} {\bibfnamefont {A.~G.}\ \bibnamefont {MacPhee}}, \bibinfo
  {author} {\bibfnamefont {J.~L.}\ \bibnamefont {Milovich}}, \bibinfo {author}
  {\bibfnamefont {J.}~\bibnamefont {Moody}}, \bibinfo {author} {\bibfnamefont
  {A.}~\bibnamefont {Pak}}, \bibinfo {author} {\bibfnamefont {H.~S.}\
  \bibnamefont {Park}}, \bibinfo {author} {\bibfnamefont {P.~K.}\ \bibnamefont
  {Patel}}, \bibinfo {author} {\bibfnamefont {J.~E.}\ \bibnamefont {Ralph}},
  \bibinfo {author} {\bibfnamefont {H.~F.}\ \bibnamefont {Robey}}, \bibinfo
  {author} {\bibfnamefont {J.~S.}\ \bibnamefont {Ross}}, \bibinfo {author}
  {\bibfnamefont {J.~D.}\ \bibnamefont {Salmonson}}, \bibinfo {author}
  {\bibfnamefont {B.~K.}\ \bibnamefont {Spears}}, \bibinfo {author}
  {\bibfnamefont {P.~T.}\ \bibnamefont {Springer}}, \bibinfo {author}
  {\bibfnamefont {R.}~\bibnamefont {Tommasini}}, \bibinfo {author}
  {\bibfnamefont {F.}~\bibnamefont {Albert}}, \bibinfo {author} {\bibfnamefont
  {L.~R.}\ \bibnamefont {Benedetti}}, \bibinfo {author} {\bibfnamefont
  {R.}~\bibnamefont {Bionta}}, \bibinfo {author} {\bibfnamefont
  {E.}~\bibnamefont {Bond}}, \bibinfo {author} {\bibfnamefont {D.~K.}\
  \bibnamefont {Bradley}}, \bibinfo {author} {\bibfnamefont {J.}~\bibnamefont
  {Caggiano}}, \bibinfo {author} {\bibfnamefont {P.~M.}\ \bibnamefont
  {Celliers}}, \bibinfo {author} {\bibfnamefont {C.}~\bibnamefont {Cerjan}},
  \bibinfo {author} {\bibfnamefont {J.~A.}\ \bibnamefont {Church}}, \bibinfo
  {author} {\bibfnamefont {R.}~\bibnamefont {Dylla-Spears}}, \bibinfo {author}
  {\bibfnamefont {D.}~\bibnamefont {Edgell}}, \bibinfo {author} {\bibfnamefont
  {M.~J.}\ \bibnamefont {Edwards}}, \bibinfo {author} {\bibfnamefont
  {D.}~\bibnamefont {Fittinghoff}}, \bibinfo {author} {\bibfnamefont {M.~A.}\
  \bibnamefont {Barrios~Garcia}}, \bibinfo {author} {\bibfnamefont
  {A.}~\bibnamefont {Hamza}}, \bibinfo {author} {\bibfnamefont
  {R.}~\bibnamefont {Hatarik}}, \bibinfo {author} {\bibfnamefont
  {H.}~\bibnamefont {Herrmann}}, \bibinfo {author} {\bibfnamefont
  {M.}~\bibnamefont {Hohenberger}}, \bibinfo {author} {\bibfnamefont
  {D.}~\bibnamefont {Hoover}}, \bibinfo {author} {\bibfnamefont {J.~L.}\
  \bibnamefont {Kline}}, \bibinfo {author} {\bibfnamefont {G.}~\bibnamefont
  {Kyrala}}, \bibinfo {author} {\bibfnamefont {B.}~\bibnamefont
  {Kozioziemski}}, \bibinfo {author} {\bibfnamefont {G.}~\bibnamefont {Grim}},
  \bibinfo {author} {\bibfnamefont {J.~E.}\ \bibnamefont {Field}}, \bibinfo
  {author} {\bibfnamefont {J.}~\bibnamefont {Frenje}}, \bibinfo {author}
  {\bibfnamefont {N.}~\bibnamefont {Izumi}}, \bibinfo {author} {\bibfnamefont
  {M.}~\bibnamefont {Gatu~Johnson}}, \bibinfo {author} {\bibfnamefont {S.~F.}\
  \bibnamefont {Khan}}, \bibinfo {author} {\bibfnamefont {J.}~\bibnamefont
  {Knauer}}, \bibinfo {author} {\bibfnamefont {T.}~\bibnamefont {Kohut}},
  \bibinfo {author} {\bibfnamefont {O.}~\bibnamefont {Landen}}, \bibinfo
  {author} {\bibfnamefont {F.}~\bibnamefont {Merrill}}, \bibinfo {author}
  {\bibfnamefont {P.}~\bibnamefont {Michel}}, \bibinfo {author} {\bibfnamefont
  {A.}~\bibnamefont {Moore}}, \bibinfo {author} {\bibfnamefont {S.~R.}\
  \bibnamefont {Nagel}}, \bibinfo {author} {\bibfnamefont {A.}~\bibnamefont
  {Nikroo}}, \bibinfo {author} {\bibfnamefont {T.}~\bibnamefont {Parham}},
  \bibinfo {author} {\bibfnamefont {R.~R.}\ \bibnamefont {Rygg}}, \bibinfo
  {author} {\bibfnamefont {D.}~\bibnamefont {Sayre}}, \bibinfo {author}
  {\bibfnamefont {M.}~\bibnamefont {Schneider}}, \bibinfo {author}
  {\bibfnamefont {D.}~\bibnamefont {Shaughnessy}}, \bibinfo {author}
  {\bibfnamefont {D.}~\bibnamefont {Strozzi}}, \bibinfo {author} {\bibfnamefont
  {R.~P.~J.}\ \bibnamefont {Town}}, \bibinfo {author} {\bibfnamefont
  {D.}~\bibnamefont {Turnbull}}, \bibinfo {author} {\bibfnamefont
  {P.}~\bibnamefont {Volegov}}, \bibinfo {author} {\bibfnamefont
  {A.}~\bibnamefont {Wan}}, \bibinfo {author} {\bibfnamefont {K.}~\bibnamefont
  {Widmann}}, \bibinfo {author} {\bibfnamefont {C.}~\bibnamefont {Wilde}}, \
  and\ \bibinfo {author} {\bibfnamefont {C.}~\bibnamefont {Yeamans}},\ }\href
  {http://dx.doi.org/10.1038/nphys3720} {\bibfield  {journal} {\bibinfo
  {journal} {Nat Phys}\ }\textbf {\bibinfo {volume} {12}},\ \bibinfo {pages}
  {800} (\bibinfo {year} {2016})}\BibitemShut {NoStop}%
\bibitem [{\citenamefont {Ross}\ \emph {et~al.}(2017)\citenamefont {Ross},
  \citenamefont {Higginson}, \citenamefont {Ryutov}, \citenamefont {Fiuza},
  \citenamefont {Hatarik}, \citenamefont {Huntington}, \citenamefont
  {Kalantar}, \citenamefont {Link}, \citenamefont {Pollock}, \citenamefont
  {Remington}, \citenamefont {Rinderknecht}, \citenamefont {Swadling},
  \citenamefont {Turnbull}, \citenamefont {Weber}, \citenamefont {Wilks},
  \citenamefont {Froula}, \citenamefont {Rosenberg}, \citenamefont {Morita},
  \citenamefont {Sakawa}, \citenamefont {Takabe}, \citenamefont {Drake},
  \citenamefont {Kuranz}, \citenamefont {Gregori}, \citenamefont {Meinecke},
  \citenamefont {Levy}, \citenamefont {Koenig}, \citenamefont {Spitkovsky},
  \citenamefont {Petrasso}, \citenamefont {Li}, \citenamefont {Sio},
  \citenamefont {Lahmann}, \citenamefont {Zylstra},\ and\ \citenamefont
  {Park}}]{Ross2017}%
  \BibitemOpen
  \bibfield  {author} {\bibinfo {author} {\bibfnamefont {J.~S.}\ \bibnamefont
  {Ross}}, \bibinfo {author} {\bibfnamefont {D.~P.}\ \bibnamefont {Higginson}},
  \bibinfo {author} {\bibfnamefont {D.}~\bibnamefont {Ryutov}}, \bibinfo
  {author} {\bibfnamefont {F.}~\bibnamefont {Fiuza}}, \bibinfo {author}
  {\bibfnamefont {R.}~\bibnamefont {Hatarik}}, \bibinfo {author} {\bibfnamefont
  {C.~M.}\ \bibnamefont {Huntington}}, \bibinfo {author} {\bibfnamefont
  {D.~H.}\ \bibnamefont {Kalantar}}, \bibinfo {author} {\bibfnamefont
  {A.}~\bibnamefont {Link}}, \bibinfo {author} {\bibfnamefont {B.~B.}\
  \bibnamefont {Pollock}}, \bibinfo {author} {\bibfnamefont {B.~A.}\
  \bibnamefont {Remington}}, \bibinfo {author} {\bibfnamefont {H.~G.}\
  \bibnamefont {Rinderknecht}}, \bibinfo {author} {\bibfnamefont {G.~F.}\
  \bibnamefont {Swadling}}, \bibinfo {author} {\bibfnamefont {D.~P.}\
  \bibnamefont {Turnbull}}, \bibinfo {author} {\bibfnamefont {S.}~\bibnamefont
  {Weber}}, \bibinfo {author} {\bibfnamefont {S.}~\bibnamefont {Wilks}},
  \bibinfo {author} {\bibfnamefont {D.~H.}\ \bibnamefont {Froula}}, \bibinfo
  {author} {\bibfnamefont {M.~J.}\ \bibnamefont {Rosenberg}}, \bibinfo {author}
  {\bibfnamefont {T.}~\bibnamefont {Morita}}, \bibinfo {author} {\bibfnamefont
  {Y.}~\bibnamefont {Sakawa}}, \bibinfo {author} {\bibfnamefont
  {H.}~\bibnamefont {Takabe}}, \bibinfo {author} {\bibfnamefont {R.~P.}\
  \bibnamefont {Drake}}, \bibinfo {author} {\bibfnamefont {C.}~\bibnamefont
  {Kuranz}}, \bibinfo {author} {\bibfnamefont {G.}~\bibnamefont {Gregori}},
  \bibinfo {author} {\bibfnamefont {J.}~\bibnamefont {Meinecke}}, \bibinfo
  {author} {\bibfnamefont {M.~C.}\ \bibnamefont {Levy}}, \bibinfo {author}
  {\bibfnamefont {M.}~\bibnamefont {Koenig}}, \bibinfo {author} {\bibfnamefont
  {A.}~\bibnamefont {Spitkovsky}}, \bibinfo {author} {\bibfnamefont {R.~D.}\
  \bibnamefont {Petrasso}}, \bibinfo {author} {\bibfnamefont {C.~K.}\
  \bibnamefont {Li}}, \bibinfo {author} {\bibfnamefont {H.}~\bibnamefont
  {Sio}}, \bibinfo {author} {\bibfnamefont {B.}~\bibnamefont {Lahmann}},
  \bibinfo {author} {\bibfnamefont {A.~B.}\ \bibnamefont {Zylstra}}, \ and\
  \bibinfo {author} {\bibfnamefont {H.-S.}\ \bibnamefont {Park}},\ }\href
  {\doibase 10.1103/PhysRevLett.118.185003} {\bibfield  {journal} {\bibinfo
  {journal} {Phys. Rev. Lett.}\ }\textbf {\bibinfo {volume} {118}},\ \bibinfo
  {pages} {185003} (\bibinfo {year} {2017})}\BibitemShut {NoStop}%
\bibitem [{\citenamefont {Clark}\ \emph {et~al.}(2000)\citenamefont {Clark},
  \citenamefont {Krushelnick}, \citenamefont {Davies}, \citenamefont {Zepf},
  \citenamefont {Tatarakis}, \citenamefont {Beg}, \citenamefont {Machacek},
  \citenamefont {Norreys}, \citenamefont {Santala}, \citenamefont {Watts},\
  and\ \citenamefont {Dangor}}]{clark2000}%
  \BibitemOpen
  \bibfield  {author} {\bibinfo {author} {\bibfnamefont {E.~L.}\ \bibnamefont
  {Clark}}, \bibinfo {author} {\bibfnamefont {K.}~\bibnamefont {Krushelnick}},
  \bibinfo {author} {\bibfnamefont {J.~R.}\ \bibnamefont {Davies}}, \bibinfo
  {author} {\bibfnamefont {M.}~\bibnamefont {Zepf}}, \bibinfo {author}
  {\bibfnamefont {M.}~\bibnamefont {Tatarakis}}, \bibinfo {author}
  {\bibfnamefont {F.~N.}\ \bibnamefont {Beg}}, \bibinfo {author} {\bibfnamefont
  {A.}~\bibnamefont {Machacek}}, \bibinfo {author} {\bibfnamefont {P.~A.}\
  \bibnamefont {Norreys}}, \bibinfo {author} {\bibfnamefont {M.~I.~K.}\
  \bibnamefont {Santala}}, \bibinfo {author} {\bibfnamefont {I.}~\bibnamefont
  {Watts}}, \ and\ \bibinfo {author} {\bibfnamefont {A.~E.}\ \bibnamefont
  {Dangor}},\ }\href {\doibase 10.1103/PhysRevLett.84.670} {\bibfield
  {journal} {\bibinfo  {journal} {Phys. Rev. Lett.}\ }\textbf {\bibinfo
  {volume} {84}},\ \bibinfo {pages} {670} (\bibinfo {year} {2000})}\BibitemShut
  {NoStop}%
\bibitem [{\citenamefont {Patel}\ \emph {et~al.}(2003)\citenamefont {Patel},
  \citenamefont {Mackinnon}, \citenamefont {Key}, \citenamefont {Cowan},
  \citenamefont {Foord}, \citenamefont {Allen}, \citenamefont {Price},
  \citenamefont {Ruhl}, \citenamefont {Springer},\ and\ \citenamefont
  {Stephens}}]{Patel2003}%
  \BibitemOpen
  \bibfield  {author} {\bibinfo {author} {\bibfnamefont {P.~K.}\ \bibnamefont
  {Patel}}, \bibinfo {author} {\bibfnamefont {A.~J.}\ \bibnamefont
  {Mackinnon}}, \bibinfo {author} {\bibfnamefont {M.~H.}\ \bibnamefont {Key}},
  \bibinfo {author} {\bibfnamefont {T.~E.}\ \bibnamefont {Cowan}}, \bibinfo
  {author} {\bibfnamefont {M.~E.}\ \bibnamefont {Foord}}, \bibinfo {author}
  {\bibfnamefont {M.}~\bibnamefont {Allen}}, \bibinfo {author} {\bibfnamefont
  {D.~F.}\ \bibnamefont {Price}}, \bibinfo {author} {\bibfnamefont
  {H.}~\bibnamefont {Ruhl}}, \bibinfo {author} {\bibfnamefont {P.~T.}\
  \bibnamefont {Springer}}, \ and\ \bibinfo {author} {\bibfnamefont
  {R.}~\bibnamefont {Stephens}},\ }\href {\doibase
  10.1103/PhysRevLett.91.125004} {\bibfield  {journal} {\bibinfo  {journal}
  {Phys. Rev. Lett.}\ }\textbf {\bibinfo {volume} {91}},\ \bibinfo {pages}
  {125004} (\bibinfo {year} {2003})}\BibitemShut {NoStop}%
\bibitem [{\citenamefont {Dyer}\ \emph {et~al.}(2008)\citenamefont {Dyer},
  \citenamefont {Bernstein}, \citenamefont {Cho}, \citenamefont {Osterholz},
  \citenamefont {Grigsby}, \citenamefont {Dalton}, \citenamefont {Shepherd},
  \citenamefont {Ping}, \citenamefont {Chen}, \citenamefont {Widmann},\ and\
  \citenamefont {Ditmire}}]{Dyer2008}%
  \BibitemOpen
  \bibfield  {author} {\bibinfo {author} {\bibfnamefont {G.~M.}\ \bibnamefont
  {Dyer}}, \bibinfo {author} {\bibfnamefont {A.~C.}\ \bibnamefont {Bernstein}},
  \bibinfo {author} {\bibfnamefont {B.~I.}\ \bibnamefont {Cho}}, \bibinfo
  {author} {\bibfnamefont {J.}~\bibnamefont {Osterholz}}, \bibinfo {author}
  {\bibfnamefont {W.}~\bibnamefont {Grigsby}}, \bibinfo {author} {\bibfnamefont
  {A.}~\bibnamefont {Dalton}}, \bibinfo {author} {\bibfnamefont
  {R.}~\bibnamefont {Shepherd}}, \bibinfo {author} {\bibfnamefont
  {Y.}~\bibnamefont {Ping}}, \bibinfo {author} {\bibfnamefont {H.}~\bibnamefont
  {Chen}}, \bibinfo {author} {\bibfnamefont {K.}~\bibnamefont {Widmann}}, \
  and\ \bibinfo {author} {\bibfnamefont {T.}~\bibnamefont {Ditmire}},\ }\href
  {\doibase 10.1103/PhysRevLett.101.015002} {\bibfield  {journal} {\bibinfo
  {journal} {Phys. Rev. Lett.}\ }\textbf {\bibinfo {volume} {101}},\ \bibinfo
  {pages} {015002} (\bibinfo {year} {2008})}\BibitemShut {NoStop}%
\bibitem [{\citenamefont {Fern\'{a}ndez}\ \emph {et~al.}(2014)\citenamefont
  {Fern\'{a}ndez}, \citenamefont {Albright}, \citenamefont {Beg}, \citenamefont
  {Foord}, \citenamefont {Hegelich}, \citenamefont {Honrubia}, \citenamefont
  {Roth}, \citenamefont {Stephens},\ and\ \citenamefont {Yin}}]{Fernandez2014}%
  \BibitemOpen
  \bibfield  {author} {\bibinfo {author} {\bibfnamefont {J.}~\bibnamefont
  {Fern\'{a}ndez}}, \bibinfo {author} {\bibfnamefont {B.}~\bibnamefont
  {Albright}}, \bibinfo {author} {\bibfnamefont {F.}~\bibnamefont {Beg}},
  \bibinfo {author} {\bibfnamefont {M.}~\bibnamefont {Foord}}, \bibinfo
  {author} {\bibfnamefont {B.}~\bibnamefont {Hegelich}}, \bibinfo {author}
  {\bibfnamefont {J.}~\bibnamefont {Honrubia}}, \bibinfo {author}
  {\bibfnamefont {M.}~\bibnamefont {Roth}}, \bibinfo {author} {\bibfnamefont
  {R.}~\bibnamefont {Stephens}}, \ and\ \bibinfo {author} {\bibfnamefont
  {L.}~\bibnamefont {Yin}},\ }\href
  {http://stacks.iop.org/0029-5515/54/i=5/a=054006} {\bibfield  {journal}
  {\bibinfo  {journal} {Nuclear Fusion}\ }\textbf {\bibinfo {volume} {54}},\
  \bibinfo {pages} {054006} (\bibinfo {year} {2014})}\BibitemShut {NoStop}%
\bibitem [{\citenamefont {Nilson}\ \emph {et~al.}(2006)\citenamefont {Nilson},
  \citenamefont {Willingale}, \citenamefont {Kaluza}, \citenamefont
  {Kamperidis}, \citenamefont {Minardi}, \citenamefont {Wei}, \citenamefont
  {Fernandes}, \citenamefont {Notley}, \citenamefont {Bandyopadhyay},
  \citenamefont {Sherlock}, \citenamefont {Kingham}, \citenamefont {Tatarakis},
  \citenamefont {Najmudin}, \citenamefont {Rozmus}, \citenamefont {Evans},
  \citenamefont {Haines}, \citenamefont {Dangor},\ and\ \citenamefont
  {Krushelnick}}]{Nilson2006}%
  \BibitemOpen
  \bibfield  {author} {\bibinfo {author} {\bibfnamefont {P.~M.}\ \bibnamefont
  {Nilson}}, \bibinfo {author} {\bibfnamefont {L.}~\bibnamefont {Willingale}},
  \bibinfo {author} {\bibfnamefont {M.~C.}\ \bibnamefont {Kaluza}}, \bibinfo
  {author} {\bibfnamefont {C.}~\bibnamefont {Kamperidis}}, \bibinfo {author}
  {\bibfnamefont {S.}~\bibnamefont {Minardi}}, \bibinfo {author} {\bibfnamefont
  {M.~S.}\ \bibnamefont {Wei}}, \bibinfo {author} {\bibfnamefont
  {P.}~\bibnamefont {Fernandes}}, \bibinfo {author} {\bibfnamefont
  {M.}~\bibnamefont {Notley}}, \bibinfo {author} {\bibfnamefont
  {S.}~\bibnamefont {Bandyopadhyay}}, \bibinfo {author} {\bibfnamefont
  {M.}~\bibnamefont {Sherlock}}, \bibinfo {author} {\bibfnamefont {R.~J.}\
  \bibnamefont {Kingham}}, \bibinfo {author} {\bibfnamefont {M.}~\bibnamefont
  {Tatarakis}}, \bibinfo {author} {\bibfnamefont {Z.}~\bibnamefont {Najmudin}},
  \bibinfo {author} {\bibfnamefont {W.}~\bibnamefont {Rozmus}}, \bibinfo
  {author} {\bibfnamefont {R.~G.}\ \bibnamefont {Evans}}, \bibinfo {author}
  {\bibfnamefont {M.~G.}\ \bibnamefont {Haines}}, \bibinfo {author}
  {\bibfnamefont {A.~E.}\ \bibnamefont {Dangor}}, \ and\ \bibinfo {author}
  {\bibfnamefont {K.}~\bibnamefont {Krushelnick}},\ }\href {\doibase
  10.1103/PhysRevLett.97.255001} {\bibfield  {journal} {\bibinfo  {journal}
  {Phys. Rev. Lett.}\ }\textbf {\bibinfo {volume} {97}},\ \bibinfo {pages}
  {255001} (\bibinfo {year} {2006})}\BibitemShut {NoStop}%
\bibitem [{\citenamefont {Snavely}\ \emph {et~al.}(2000)\citenamefont
  {Snavely}, \citenamefont {Key}, \citenamefont {Hatchett}, \citenamefont
  {Cowan}, \citenamefont {Roth}, \citenamefont {Phillips}, \citenamefont
  {Stoyer}, \citenamefont {Henry}, \citenamefont {Sangster}, \citenamefont
  {Singh}, \citenamefont {Wilks}, \citenamefont {MacKinnon}, \citenamefont
  {Offenberger}, \citenamefont {Pennington}, \citenamefont {Yasuike},
  \citenamefont {Langdon}, \citenamefont {Lasinski}, \citenamefont {Johnson},
  \citenamefont {Perry},\ and\ \citenamefont {Campbell}}]{Snavely2000}%
  \BibitemOpen
  \bibfield  {author} {\bibinfo {author} {\bibfnamefont {R.~A.}\ \bibnamefont
  {Snavely}}, \bibinfo {author} {\bibfnamefont {M.~H.}\ \bibnamefont {Key}},
  \bibinfo {author} {\bibfnamefont {S.~P.}\ \bibnamefont {Hatchett}}, \bibinfo
  {author} {\bibfnamefont {T.~E.}\ \bibnamefont {Cowan}}, \bibinfo {author}
  {\bibfnamefont {M.}~\bibnamefont {Roth}}, \bibinfo {author} {\bibfnamefont
  {T.~W.}\ \bibnamefont {Phillips}}, \bibinfo {author} {\bibfnamefont {M.~A.}\
  \bibnamefont {Stoyer}}, \bibinfo {author} {\bibfnamefont {E.~A.}\
  \bibnamefont {Henry}}, \bibinfo {author} {\bibfnamefont {T.~C.}\ \bibnamefont
  {Sangster}}, \bibinfo {author} {\bibfnamefont {M.~S.}\ \bibnamefont {Singh}},
  \bibinfo {author} {\bibfnamefont {S.~C.}\ \bibnamefont {Wilks}}, \bibinfo
  {author} {\bibfnamefont {A.}~\bibnamefont {MacKinnon}}, \bibinfo {author}
  {\bibfnamefont {A.}~\bibnamefont {Offenberger}}, \bibinfo {author}
  {\bibfnamefont {D.~M.}\ \bibnamefont {Pennington}}, \bibinfo {author}
  {\bibfnamefont {K.}~\bibnamefont {Yasuike}}, \bibinfo {author} {\bibfnamefont
  {A.~B.}\ \bibnamefont {Langdon}}, \bibinfo {author} {\bibfnamefont {B.~F.}\
  \bibnamefont {Lasinski}}, \bibinfo {author} {\bibfnamefont {J.}~\bibnamefont
  {Johnson}}, \bibinfo {author} {\bibfnamefont {M.~D.}\ \bibnamefont {Perry}},
  \ and\ \bibinfo {author} {\bibfnamefont {E.~M.}\ \bibnamefont {Campbell}},\
  }\href {\doibase 10.1103/PhysRevLett.85.2945} {\bibfield  {journal} {\bibinfo
   {journal} {Phys. Rev. Lett.}\ }\textbf {\bibinfo {volume} {85}},\ \bibinfo
  {pages} {2945} (\bibinfo {year} {2000})}\BibitemShut {NoStop}%
\bibitem [{\citenamefont {Tikhonchuk}\ \emph {et~al.}(2005)\citenamefont
  {Tikhonchuk}, \citenamefont {Andreev}, \citenamefont {Bochkarev},\ and\
  \citenamefont {Bychenkov}}]{Tikhonchuk2005}%
  \BibitemOpen
  \bibfield  {author} {\bibinfo {author} {\bibfnamefont {V.~T.}\ \bibnamefont
  {Tikhonchuk}}, \bibinfo {author} {\bibfnamefont {A.~A.}\ \bibnamefont
  {Andreev}}, \bibinfo {author} {\bibfnamefont {S.~G.}\ \bibnamefont
  {Bochkarev}}, \ and\ \bibinfo {author} {\bibfnamefont {V.~Y.}\ \bibnamefont
  {Bychenkov}},\ }\href {http://stacks.iop.org/0741-3335/47/i=12B/a=S69}
  {\bibfield  {journal} {\bibinfo  {journal} {Plasma Physics and Controlled
  Fusion}\ }\textbf {\bibinfo {volume} {47}},\ \bibinfo {pages} {B869}
  (\bibinfo {year} {2005})}\BibitemShut {NoStop}%
\bibitem [{\citenamefont {Bulanov}\ \emph {et~al.}(2008)\citenamefont
  {Bulanov}, \citenamefont {Brantov}, \citenamefont {Bychenkov}, \citenamefont
  {Chvykov}, \citenamefont {Kalinchenko}, \citenamefont {Matsuoka},
  \citenamefont {Rousseau}, \citenamefont {Reed}, \citenamefont {Yanovsky},
  \citenamefont {Litzenberg}, \citenamefont {Krushelnick},\ and\ \citenamefont
  {Maksimchuk}}]{Bulanov2008}%
  \BibitemOpen
  \bibfield  {author} {\bibinfo {author} {\bibfnamefont {S.~S.}\ \bibnamefont
  {Bulanov}}, \bibinfo {author} {\bibfnamefont {A.}~\bibnamefont {Brantov}},
  \bibinfo {author} {\bibfnamefont {V.~Y.}\ \bibnamefont {Bychenkov}}, \bibinfo
  {author} {\bibfnamefont {V.}~\bibnamefont {Chvykov}}, \bibinfo {author}
  {\bibfnamefont {G.}~\bibnamefont {Kalinchenko}}, \bibinfo {author}
  {\bibfnamefont {T.}~\bibnamefont {Matsuoka}}, \bibinfo {author}
  {\bibfnamefont {P.}~\bibnamefont {Rousseau}}, \bibinfo {author}
  {\bibfnamefont {S.}~\bibnamefont {Reed}}, \bibinfo {author} {\bibfnamefont
  {V.}~\bibnamefont {Yanovsky}}, \bibinfo {author} {\bibfnamefont {D.~W.}\
  \bibnamefont {Litzenberg}}, \bibinfo {author} {\bibfnamefont
  {K.}~\bibnamefont {Krushelnick}}, \ and\ \bibinfo {author} {\bibfnamefont
  {A.}~\bibnamefont {Maksimchuk}},\ }\href {\doibase
  10.1103/PhysRevE.78.026412} {\bibfield  {journal} {\bibinfo  {journal} {Phys.
  Rev. E}\ }\textbf {\bibinfo {volume} {78}},\ \bibinfo {pages} {026412}
  (\bibinfo {year} {2008})}\BibitemShut {NoStop}%
\bibitem [{\citenamefont {Esirkepov}\ \emph {et~al.}(2004)\citenamefont
  {Esirkepov}, \citenamefont {Borghesi}, \citenamefont {Bulanov}, \citenamefont
  {Mourou},\ and\ \citenamefont {Tajima}}]{Esirkepov2004}%
  \BibitemOpen
  \bibfield  {author} {\bibinfo {author} {\bibfnamefont {T.}~\bibnamefont
  {Esirkepov}}, \bibinfo {author} {\bibfnamefont {M.}~\bibnamefont {Borghesi}},
  \bibinfo {author} {\bibfnamefont {S.~V.}\ \bibnamefont {Bulanov}}, \bibinfo
  {author} {\bibfnamefont {G.}~\bibnamefont {Mourou}}, \ and\ \bibinfo {author}
  {\bibfnamefont {T.}~\bibnamefont {Tajima}},\ }\href {\doibase
  10.1103/PhysRevLett.92.175003} {\bibfield  {journal} {\bibinfo  {journal}
  {Phys. Rev. Lett.}\ }\textbf {\bibinfo {volume} {92}},\ \bibinfo {pages}
  {175003} (\bibinfo {year} {2004})}\BibitemShut {NoStop}%
\bibitem [{\citenamefont {Hegelich}\ \emph {et~al.}(2006)\citenamefont
  {Hegelich}, \citenamefont {Albright}, \citenamefont {Cobble}, \citenamefont
  {Flippo}, \citenamefont {Letzring}, \citenamefont {Paffett}, \citenamefont
  {Ruhl}, \citenamefont {Schreiber}, \citenamefont {Schulze},\ and\
  \citenamefont {Fern\'{a}ndez}}]{Hegelich2005}%
  \BibitemOpen
  \bibfield  {author} {\bibinfo {author} {\bibfnamefont {B.~M.}\ \bibnamefont
  {Hegelich}}, \bibinfo {author} {\bibfnamefont {B.~J.}\ \bibnamefont
  {Albright}}, \bibinfo {author} {\bibfnamefont {J.}~\bibnamefont {Cobble}},
  \bibinfo {author} {\bibfnamefont {K.}~\bibnamefont {Flippo}}, \bibinfo
  {author} {\bibfnamefont {S.}~\bibnamefont {Letzring}}, \bibinfo {author}
  {\bibfnamefont {M.}~\bibnamefont {Paffett}}, \bibinfo {author} {\bibfnamefont
  {H.}~\bibnamefont {Ruhl}}, \bibinfo {author} {\bibfnamefont {J.}~\bibnamefont
  {Schreiber}}, \bibinfo {author} {\bibfnamefont {R.~K.}\ \bibnamefont
  {Schulze}}, \ and\ \bibinfo {author} {\bibfnamefont {J.~C.}\ \bibnamefont
  {Fern\'{a}ndez}},\ }\href {http://dx.doi.org/10.1038/nature04400} {\bibfield
  {journal} {\bibinfo  {journal} {Nature}\ }\textbf {\bibinfo {volume} {439}},\
  \bibinfo {pages} {441} (\bibinfo {year} {2006})}\BibitemShut {NoStop}%
\bibitem [{\citenamefont {Palmer}\ \emph {et~al.}(2011)\citenamefont {Palmer},
  \citenamefont {Dover}, \citenamefont {Pogorelsky}, \citenamefont {Babzien},
  \citenamefont {Dudnikova}, \citenamefont {Ispiriyan}, \citenamefont
  {Polyanskiy}, \citenamefont {Schreiber}, \citenamefont {Shkolnikov},
  \citenamefont {Yakimenko},\ and\ \citenamefont {Najmudin}}]{Palmer2011}%
  \BibitemOpen
  \bibfield  {author} {\bibinfo {author} {\bibfnamefont {C.~A.~J.}\
  \bibnamefont {Palmer}}, \bibinfo {author} {\bibfnamefont {N.~P.}\
  \bibnamefont {Dover}}, \bibinfo {author} {\bibfnamefont {I.}~\bibnamefont
  {Pogorelsky}}, \bibinfo {author} {\bibfnamefont {M.}~\bibnamefont {Babzien}},
  \bibinfo {author} {\bibfnamefont {G.~I.}\ \bibnamefont {Dudnikova}}, \bibinfo
  {author} {\bibfnamefont {M.}~\bibnamefont {Ispiriyan}}, \bibinfo {author}
  {\bibfnamefont {M.~N.}\ \bibnamefont {Polyanskiy}}, \bibinfo {author}
  {\bibfnamefont {J.}~\bibnamefont {Schreiber}}, \bibinfo {author}
  {\bibfnamefont {P.}~\bibnamefont {Shkolnikov}}, \bibinfo {author}
  {\bibfnamefont {V.}~\bibnamefont {Yakimenko}}, \ and\ \bibinfo {author}
  {\bibfnamefont {Z.}~\bibnamefont {Najmudin}},\ }\href {\doibase
  10.1103/PhysRevLett.106.014801} {\bibfield  {journal} {\bibinfo  {journal}
  {Phys. Rev. Lett.}\ }\textbf {\bibinfo {volume} {106}},\ \bibinfo {pages}
  {014801} (\bibinfo {year} {2011})}\BibitemShut {NoStop}%
\bibitem [{\citenamefont {Henig}\ \emph {et~al.}(2009)\citenamefont {Henig},
  \citenamefont {Steinke}, \citenamefont {Schn\"urer}, \citenamefont
  {Sokollik}, \citenamefont {H\"orlein}, \citenamefont {Kiefer}, \citenamefont
  {Jung}, \citenamefont {Schreiber}, \citenamefont {Hegelich}, \citenamefont
  {Yan}, \citenamefont {Meyer-ter Vehn}, \citenamefont {Tajima}, \citenamefont
  {Nickles}, \citenamefont {Sandner},\ and\ \citenamefont {Habs}}]{Henig2009}%
  \BibitemOpen
  \bibfield  {author} {\bibinfo {author} {\bibfnamefont {A.}~\bibnamefont
  {Henig}}, \bibinfo {author} {\bibfnamefont {S.}~\bibnamefont {Steinke}},
  \bibinfo {author} {\bibfnamefont {M.}~\bibnamefont {Schn\"urer}}, \bibinfo
  {author} {\bibfnamefont {T.}~\bibnamefont {Sokollik}}, \bibinfo {author}
  {\bibfnamefont {R.}~\bibnamefont {H\"orlein}}, \bibinfo {author}
  {\bibfnamefont {D.}~\bibnamefont {Kiefer}}, \bibinfo {author} {\bibfnamefont
  {D.}~\bibnamefont {Jung}}, \bibinfo {author} {\bibfnamefont {J.}~\bibnamefont
  {Schreiber}}, \bibinfo {author} {\bibfnamefont {B.~M.}\ \bibnamefont
  {Hegelich}}, \bibinfo {author} {\bibfnamefont {X.~Q.}\ \bibnamefont {Yan}},
  \bibinfo {author} {\bibfnamefont {J.}~\bibnamefont {Meyer-ter Vehn}},
  \bibinfo {author} {\bibfnamefont {T.}~\bibnamefont {Tajima}}, \bibinfo
  {author} {\bibfnamefont {P.~V.}\ \bibnamefont {Nickles}}, \bibinfo {author}
  {\bibfnamefont {W.}~\bibnamefont {Sandner}}, \ and\ \bibinfo {author}
  {\bibfnamefont {D.}~\bibnamefont {Habs}},\ }\href {\doibase
  10.1103/PhysRevLett.103.245003} {\bibfield  {journal} {\bibinfo  {journal}
  {Phys. Rev. Lett.}\ }\textbf {\bibinfo {volume} {103}},\ \bibinfo {pages}
  {245003} (\bibinfo {year} {2009})}\BibitemShut {NoStop}%
\bibitem [{\citenamefont {Kar}\ \emph {et~al.}(2012)\citenamefont {Kar},
  \citenamefont {Kakolee}, \citenamefont {Qiao}, \citenamefont {Macchi},
  \citenamefont {Cerchez}, \citenamefont {Doria}, \citenamefont {Geissler},
  \citenamefont {McKenna}, \citenamefont {Neely}, \citenamefont {Osterholz},
  \citenamefont {Prasad}, \citenamefont {Quinn}, \citenamefont {Ramakrishna},
  \citenamefont {Sarri}, \citenamefont {Willi}, \citenamefont {Yuan},
  \citenamefont {Zepf},\ and\ \citenamefont {Borghesi}}]{Kar2012}%
  \BibitemOpen
  \bibfield  {author} {\bibinfo {author} {\bibfnamefont {S.}~\bibnamefont
  {Kar}}, \bibinfo {author} {\bibfnamefont {K.~F.}\ \bibnamefont {Kakolee}},
  \bibinfo {author} {\bibfnamefont {B.}~\bibnamefont {Qiao}}, \bibinfo {author}
  {\bibfnamefont {A.}~\bibnamefont {Macchi}}, \bibinfo {author} {\bibfnamefont
  {M.}~\bibnamefont {Cerchez}}, \bibinfo {author} {\bibfnamefont
  {D.}~\bibnamefont {Doria}}, \bibinfo {author} {\bibfnamefont
  {M.}~\bibnamefont {Geissler}}, \bibinfo {author} {\bibfnamefont
  {P.}~\bibnamefont {McKenna}}, \bibinfo {author} {\bibfnamefont
  {D.}~\bibnamefont {Neely}}, \bibinfo {author} {\bibfnamefont
  {J.}~\bibnamefont {Osterholz}}, \bibinfo {author} {\bibfnamefont
  {R.}~\bibnamefont {Prasad}}, \bibinfo {author} {\bibfnamefont
  {K.}~\bibnamefont {Quinn}}, \bibinfo {author} {\bibfnamefont
  {B.}~\bibnamefont {Ramakrishna}}, \bibinfo {author} {\bibfnamefont
  {G.}~\bibnamefont {Sarri}}, \bibinfo {author} {\bibfnamefont
  {O.}~\bibnamefont {Willi}}, \bibinfo {author} {\bibfnamefont {X.~Y.}\
  \bibnamefont {Yuan}}, \bibinfo {author} {\bibfnamefont {M.}~\bibnamefont
  {Zepf}}, \ and\ \bibinfo {author} {\bibfnamefont {M.}~\bibnamefont
  {Borghesi}},\ }\href {\doibase 10.1103/PhysRevLett.109.185006} {\bibfield
  {journal} {\bibinfo  {journal} {Phys. Rev. Lett.}\ }\textbf {\bibinfo
  {volume} {109}},\ \bibinfo {pages} {185006} (\bibinfo {year}
  {2012})}\BibitemShut {NoStop}%
\bibitem [{\citenamefont {Palaniyappan}\ \emph {et~al.}(2015)\citenamefont
  {Palaniyappan}, \citenamefont {Huang}, \citenamefont {Gautier}, \citenamefont
  {Hamilton}, \citenamefont {Santiago}, \citenamefont {Kreuzer}, \citenamefont
  {Sefkow}, \citenamefont {Shah},\ and\ \citenamefont
  {Fern{\'a}ndez}}]{Palaniyappan2015}%
  \BibitemOpen
  \bibfield  {author} {\bibinfo {author} {\bibfnamefont {S.}~\bibnamefont
  {Palaniyappan}}, \bibinfo {author} {\bibfnamefont {C.}~\bibnamefont {Huang}},
  \bibinfo {author} {\bibfnamefont {D.~C.}\ \bibnamefont {Gautier}}, \bibinfo
  {author} {\bibfnamefont {C.~E.}\ \bibnamefont {Hamilton}}, \bibinfo {author}
  {\bibfnamefont {M.~A.}\ \bibnamefont {Santiago}}, \bibinfo {author}
  {\bibfnamefont {C.}~\bibnamefont {Kreuzer}}, \bibinfo {author} {\bibfnamefont
  {A.~B.}\ \bibnamefont {Sefkow}}, \bibinfo {author} {\bibfnamefont {R.~C.}\
  \bibnamefont {Shah}}, \ and\ \bibinfo {author} {\bibfnamefont {J.~C.}\
  \bibnamefont {Fern{\'a}ndez}},\ }\href
  {http://dx.doi.org/10.1038/ncomms10170} {\ \textbf {\bibinfo {volume} {6}},\
  \bibinfo {pages} {10170 EP } (\bibinfo {year} {2015})}\BibitemShut {NoStop}%
\bibitem [{\citenamefont {Haberberger}\ \emph {et~al.}(2012)\citenamefont
  {Haberberger}, \citenamefont {Tochitsky}, \citenamefont {Fiuza},
  \citenamefont {Gong}, \citenamefont {Fonseca}, \citenamefont {Silva},
  \citenamefont {Mori},\ and\ \citenamefont {Joshi}}]{Haberberger2012}%
  \BibitemOpen
  \bibfield  {author} {\bibinfo {author} {\bibfnamefont {D.}~\bibnamefont
  {Haberberger}}, \bibinfo {author} {\bibfnamefont {S.}~\bibnamefont
  {Tochitsky}}, \bibinfo {author} {\bibfnamefont {F.}~\bibnamefont {Fiuza}},
  \bibinfo {author} {\bibfnamefont {C.}~\bibnamefont {Gong}}, \bibinfo {author}
  {\bibfnamefont {R.~A.}\ \bibnamefont {Fonseca}}, \bibinfo {author}
  {\bibfnamefont {L.~O.}\ \bibnamefont {Silva}}, \bibinfo {author}
  {\bibfnamefont {W.~B.}\ \bibnamefont {Mori}}, \ and\ \bibinfo {author}
  {\bibfnamefont {C.}~\bibnamefont {Joshi}},\ }\href
  {http://dx.doi.org/10.1038/nphys2130} {\bibfield  {journal} {\bibinfo
  {journal} {Nat Phys}\ }\textbf {\bibinfo {volume} {8}},\ \bibinfo {pages}
  {95} (\bibinfo {year} {2012})}\BibitemShut {NoStop}%
\bibitem [{\citenamefont {Tresca}\ \emph {et~al.}(2015)\citenamefont {Tresca},
  \citenamefont {Dover}, \citenamefont {Cook}, \citenamefont {Maharjan},
  \citenamefont {Polyanskiy}, \citenamefont {Najmudin}, \citenamefont
  {Shkolnikov},\ and\ \citenamefont {Pogorelsky}}]{Tresca2015}%
  \BibitemOpen
  \bibfield  {author} {\bibinfo {author} {\bibfnamefont {O.}~\bibnamefont
  {Tresca}}, \bibinfo {author} {\bibfnamefont {N.~P.}\ \bibnamefont {Dover}},
  \bibinfo {author} {\bibfnamefont {N.}~\bibnamefont {Cook}}, \bibinfo {author}
  {\bibfnamefont {C.}~\bibnamefont {Maharjan}}, \bibinfo {author}
  {\bibfnamefont {M.~N.}\ \bibnamefont {Polyanskiy}}, \bibinfo {author}
  {\bibfnamefont {Z.}~\bibnamefont {Najmudin}}, \bibinfo {author}
  {\bibfnamefont {P.}~\bibnamefont {Shkolnikov}}, \ and\ \bibinfo {author}
  {\bibfnamefont {I.}~\bibnamefont {Pogorelsky}},\ }\href {\doibase
  10.1103/PhysRevLett.115.094802} {\bibfield  {journal} {\bibinfo  {journal}
  {Phys. Rev. Lett.}\ }\textbf {\bibinfo {volume} {115}},\ \bibinfo {pages}
  {094802} (\bibinfo {year} {2015})}\BibitemShut {NoStop}%
\bibitem [{\citenamefont {Silva}\ \emph {et~al.}(2004)\citenamefont {Silva},
  \citenamefont {Marti}, \citenamefont {Davies}, \citenamefont {Fonseca},
  \citenamefont {Ren}, \citenamefont {Tsung},\ and\ \citenamefont
  {Mori}}]{Silva2004}%
  \BibitemOpen
  \bibfield  {author} {\bibinfo {author} {\bibfnamefont {L.~O.}\ \bibnamefont
  {Silva}}, \bibinfo {author} {\bibfnamefont {M.}~\bibnamefont {Marti}},
  \bibinfo {author} {\bibfnamefont {J.~R.}\ \bibnamefont {Davies}}, \bibinfo
  {author} {\bibfnamefont {R.~A.}\ \bibnamefont {Fonseca}}, \bibinfo {author}
  {\bibfnamefont {C.}~\bibnamefont {Ren}}, \bibinfo {author} {\bibfnamefont
  {F.~S.}\ \bibnamefont {Tsung}}, \ and\ \bibinfo {author} {\bibfnamefont
  {W.~B.}\ \bibnamefont {Mori}},\ }\href {\doibase
  10.1103/PhysRevLett.92.015002} {\bibfield  {journal} {\bibinfo  {journal}
  {Phys. Rev. Lett.}\ }\textbf {\bibinfo {volume} {92}},\ \bibinfo {pages}
  {015002} (\bibinfo {year} {2004})}\BibitemShut {NoStop}%
\bibitem [{\citenamefont {Fiuza}\ \emph {et~al.}(2012)\citenamefont {Fiuza},
  \citenamefont {Stockem}, \citenamefont {Boella}, \citenamefont {Fonseca},
  \citenamefont {Silva}, \citenamefont {Haberberger}, \citenamefont
  {Tochitsky}, \citenamefont {Gong}, \citenamefont {Mori},\ and\ \citenamefont
  {Joshi}}]{Fiuza2012}%
  \BibitemOpen
  \bibfield  {author} {\bibinfo {author} {\bibfnamefont {F.}~\bibnamefont
  {Fiuza}}, \bibinfo {author} {\bibfnamefont {A.}~\bibnamefont {Stockem}},
  \bibinfo {author} {\bibfnamefont {E.}~\bibnamefont {Boella}}, \bibinfo
  {author} {\bibfnamefont {R.~A.}\ \bibnamefont {Fonseca}}, \bibinfo {author}
  {\bibfnamefont {L.~O.}\ \bibnamefont {Silva}}, \bibinfo {author}
  {\bibfnamefont {D.}~\bibnamefont {Haberberger}}, \bibinfo {author}
  {\bibfnamefont {S.}~\bibnamefont {Tochitsky}}, \bibinfo {author}
  {\bibfnamefont {C.}~\bibnamefont {Gong}}, \bibinfo {author} {\bibfnamefont
  {W.~B.}\ \bibnamefont {Mori}}, \ and\ \bibinfo {author} {\bibfnamefont
  {C.}~\bibnamefont {Joshi}},\ }\href {\doibase 10.1103/PhysRevLett.109.215001}
  {\bibfield  {journal} {\bibinfo  {journal} {Phys. Rev. Lett.}\ }\textbf
  {\bibinfo {volume} {109}},\ \bibinfo {pages} {215001} (\bibinfo {year}
  {2012})}\BibitemShut {NoStop}%
\bibitem [{\citenamefont {Medvedev}(2014)}]{Medvedev2014}%
  \BibitemOpen
  \bibfield  {author} {\bibinfo {author} {\bibfnamefont {Y.~V.}\ \bibnamefont
  {Medvedev}},\ }\href {http://stacks.iop.org/0741-3335/56/i=2/a=025005}
  {\bibfield  {journal} {\bibinfo  {journal} {Plasma Physics and Controlled
  Fusion}\ }\textbf {\bibinfo {volume} {56}},\ \bibinfo {pages} {025005}
  (\bibinfo {year} {2014})}\BibitemShut {NoStop}%
\bibitem [{\citenamefont {Gong}\ \emph {et~al.}(2016)\citenamefont {Gong},
  \citenamefont {Tochitsky}, \citenamefont {Fiuza}, \citenamefont {Pigeon},\
  and\ \citenamefont {Joshi}}]{Gong2016}%
  \BibitemOpen
  \bibfield  {author} {\bibinfo {author} {\bibfnamefont {C.}~\bibnamefont
  {Gong}}, \bibinfo {author} {\bibfnamefont {S.~Y.}\ \bibnamefont {Tochitsky}},
  \bibinfo {author} {\bibfnamefont {F.}~\bibnamefont {Fiuza}}, \bibinfo
  {author} {\bibfnamefont {J.~J.}\ \bibnamefont {Pigeon}}, \ and\ \bibinfo
  {author} {\bibfnamefont {C.}~\bibnamefont {Joshi}},\ }\href {\doibase
  10.1103/PhysRevE.93.061202} {\bibfield  {journal} {\bibinfo  {journal} {Phys.
  Rev. E}\ }\textbf {\bibinfo {volume} {93}},\ \bibinfo {pages} {061202}
  (\bibinfo {year} {2016})}\BibitemShut {NoStop}%
\bibitem [{\citenamefont {Grismayer}\ \emph {et~al.}(2008)\citenamefont
  {Grismayer}, \citenamefont {Mora}, \citenamefont {Adam},\ and\ \citenamefont
  {H\'eron}}]{Grismayer2008}%
  \BibitemOpen
  \bibfield  {author} {\bibinfo {author} {\bibfnamefont {T.}~\bibnamefont
  {Grismayer}}, \bibinfo {author} {\bibfnamefont {P.}~\bibnamefont {Mora}},
  \bibinfo {author} {\bibfnamefont {J.~C.}\ \bibnamefont {Adam}}, \ and\
  \bibinfo {author} {\bibfnamefont {A.}~\bibnamefont {H\'eron}},\ }\href
  {\doibase 10.1103/PhysRevE.77.066407} {\bibfield  {journal} {\bibinfo
  {journal} {Phys. Rev. E}\ }\textbf {\bibinfo {volume} {77}},\ \bibinfo
  {pages} {066407} (\bibinfo {year} {2008})}\BibitemShut {NoStop}%
\bibitem [{\citenamefont {Fiuza}\ \emph {et~al.}(2013)\citenamefont {Fiuza},
  \citenamefont {Stockem}, \citenamefont {Boella}, \citenamefont {Fonseca},
  \citenamefont {Silva}, \citenamefont {Haberberger}, \citenamefont
  {Tochitsky}, \citenamefont {Mori},\ and\ \citenamefont {Joshi}}]{Fiuza2013}%
  \BibitemOpen
  \bibfield  {author} {\bibinfo {author} {\bibfnamefont {F.}~\bibnamefont
  {Fiuza}}, \bibinfo {author} {\bibfnamefont {A.}~\bibnamefont {Stockem}},
  \bibinfo {author} {\bibfnamefont {E.}~\bibnamefont {Boella}}, \bibinfo
  {author} {\bibfnamefont {R.~A.}\ \bibnamefont {Fonseca}}, \bibinfo {author}
  {\bibfnamefont {L.~O.}\ \bibnamefont {Silva}}, \bibinfo {author}
  {\bibfnamefont {D.}~\bibnamefont {Haberberger}}, \bibinfo {author}
  {\bibfnamefont {S.}~\bibnamefont {Tochitsky}}, \bibinfo {author}
  {\bibfnamefont {W.~B.}\ \bibnamefont {Mori}}, \ and\ \bibinfo {author}
  {\bibfnamefont {C.}~\bibnamefont {Joshi}},\ }\href {\doibase
  10.1063/1.4801526} {\bibfield  {journal} {\bibinfo  {journal} {Physics of
  Plasmas}\ }\textbf {\bibinfo {volume} {20}},\ \bibinfo {pages} {056304}
  (\bibinfo {year} {2013})}\BibitemShut {NoStop}%
\bibitem [{Sup()}]{Supp}%
  \BibitemOpen
  \href@noop {} {\bibinfo  {journal} {See supplemental material}\ }\BibitemShut
  {NoStop}%
\bibitem [{\citenamefont {Marinak}\ \emph {et~al.}(2001)\citenamefont
  {Marinak}, \citenamefont {Kerbel}, \citenamefont {Gentile}, \citenamefont
  {Jones}, \citenamefont {Munro}, \citenamefont {Pollaine}, \citenamefont
  {Dittrich},\ and\ \citenamefont {Haan}}]{HYDRA}%
  \BibitemOpen
\bibfield  {journal} {  }\bibfield  {author} {\bibinfo {author} {\bibfnamefont
  {M.~M.}\ \bibnamefont {Marinak}}, \bibinfo {author} {\bibfnamefont {G.~D.}\
  \bibnamefont {Kerbel}}, \bibinfo {author} {\bibfnamefont {N.~A.}\
  \bibnamefont {Gentile}}, \bibinfo {author} {\bibfnamefont {O.}~\bibnamefont
  {Jones}}, \bibinfo {author} {\bibfnamefont {D.}~\bibnamefont {Munro}},
  \bibinfo {author} {\bibfnamefont {S.}~\bibnamefont {Pollaine}}, \bibinfo
  {author} {\bibfnamefont {T.~R.}\ \bibnamefont {Dittrich}}, \ and\ \bibinfo
  {author} {\bibfnamefont {S.~W.}\ \bibnamefont {Haan}},\ }\href@noop {}
  {\bibfield  {journal} {\bibinfo  {journal} {Phys. Plasmas}\ }\textbf
  {\bibinfo {volume} {8}},\ \bibinfo {pages} {2275} (\bibinfo {year}
  {2001})}\BibitemShut {NoStop}%
\bibitem [{\citenamefont {Chen}\ \emph {et~al.}(2010)\citenamefont {Chen},
  \citenamefont {Hazi}, \citenamefont {van Maren}, \citenamefont {Chen},
  \citenamefont {Fuchs}, \citenamefont {Gauthier}, \citenamefont {Pape},
  \citenamefont {Rygg},\ and\ \citenamefont {Shepherd}}]{Chen2010}%
  \BibitemOpen
  \bibfield  {author} {\bibinfo {author} {\bibfnamefont {H.}~\bibnamefont
  {Chen}}, \bibinfo {author} {\bibfnamefont {A.~U.}\ \bibnamefont {Hazi}},
  \bibinfo {author} {\bibfnamefont {R.}~\bibnamefont {van Maren}}, \bibinfo
  {author} {\bibfnamefont {S.~N.}\ \bibnamefont {Chen}}, \bibinfo {author}
  {\bibfnamefont {J.}~\bibnamefont {Fuchs}}, \bibinfo {author} {\bibfnamefont
  {M.}~\bibnamefont {Gauthier}}, \bibinfo {author} {\bibfnamefont {S.~L.}\
  \bibnamefont {Pape}}, \bibinfo {author} {\bibfnamefont {J.~R.}\ \bibnamefont
  {Rygg}}, \ and\ \bibinfo {author} {\bibfnamefont {R.}~\bibnamefont
  {Shepherd}},\ }\href {\doibase 10.1063/1.3483212} {\bibfield  {journal}
  {\bibinfo  {journal} {Review of Scientific Instruments}\ }\textbf {\bibinfo
  {volume} {81}},\ \bibinfo {pages} {10D314} (\bibinfo {year}
  {2010})}\BibitemShut {NoStop}%
\bibitem [{\citenamefont {Fuchs}\ \emph {et~al.}(2005)\citenamefont {Fuchs},
  \citenamefont {Antici}, \citenamefont {d'Humi{\`e}res}, \citenamefont
  {Lefebvre}, \citenamefont {Borghesi}, \citenamefont {Brambrink},
  \citenamefont {Cecchetti}, \citenamefont {Kaluza}, \citenamefont {Malka},
  \citenamefont {Manclossi}, \citenamefont {Meyroneinc}, \citenamefont {Mora},
  \citenamefont {Schreiber}, \citenamefont {Toncian}, \citenamefont
  {P{\'e}pin},\ and\ \citenamefont {Audebert}}]{Fuchs2005}%
  \BibitemOpen
  \bibfield  {author} {\bibinfo {author} {\bibfnamefont {J.}~\bibnamefont
  {Fuchs}}, \bibinfo {author} {\bibfnamefont {P.}~\bibnamefont {Antici}},
  \bibinfo {author} {\bibfnamefont {E.}~\bibnamefont {d'Humi{\`e}res}},
  \bibinfo {author} {\bibfnamefont {E.}~\bibnamefont {Lefebvre}}, \bibinfo
  {author} {\bibfnamefont {M.}~\bibnamefont {Borghesi}}, \bibinfo {author}
  {\bibfnamefont {E.}~\bibnamefont {Brambrink}}, \bibinfo {author}
  {\bibfnamefont {C.~A.}\ \bibnamefont {Cecchetti}}, \bibinfo {author}
  {\bibfnamefont {M.}~\bibnamefont {Kaluza}}, \bibinfo {author} {\bibfnamefont
  {V.}~\bibnamefont {Malka}}, \bibinfo {author} {\bibfnamefont
  {M.}~\bibnamefont {Manclossi}}, \bibinfo {author} {\bibfnamefont
  {S.}~\bibnamefont {Meyroneinc}}, \bibinfo {author} {\bibfnamefont
  {P.}~\bibnamefont {Mora}}, \bibinfo {author} {\bibfnamefont {J.}~\bibnamefont
  {Schreiber}}, \bibinfo {author} {\bibfnamefont {T.}~\bibnamefont {Toncian}},
  \bibinfo {author} {\bibfnamefont {H.}~\bibnamefont {P{\'e}pin}}, \ and\
  \bibinfo {author} {\bibfnamefont {P.}~\bibnamefont {Audebert}},\ }\href
  {http://dx.doi.org/10.1038/nphys199} {\bibfield  {journal} {\bibinfo
  {journal} {Nat. Phys.}\ }\textbf {\bibinfo {volume} {2}},\ \bibinfo {pages}
  {48} (\bibinfo {year} {2005})}\BibitemShut {NoStop}%
\bibitem [{\citenamefont {Toncian}\ \emph {et~al.}(2006)\citenamefont
  {Toncian}, \citenamefont {Borghesi}, \citenamefont {Fuchs}, \citenamefont
  {d{\textquoteright}Humi{\`e}res}, \citenamefont {Antici}, \citenamefont
  {Audebert}, \citenamefont {Brambrink}, \citenamefont {Cecchetti},
  \citenamefont {Pipahl}, \citenamefont {Romagnani},\ and\ \citenamefont
  {Willi}}]{Toncian2006}%
  \BibitemOpen
  \bibfield  {author} {\bibinfo {author} {\bibfnamefont {T.}~\bibnamefont
  {Toncian}}, \bibinfo {author} {\bibfnamefont {M.}~\bibnamefont {Borghesi}},
  \bibinfo {author} {\bibfnamefont {J.}~\bibnamefont {Fuchs}}, \bibinfo
  {author} {\bibfnamefont {E.}~\bibnamefont {d{\textquoteright}Humi{\`e}res}},
  \bibinfo {author} {\bibfnamefont {P.}~\bibnamefont {Antici}}, \bibinfo
  {author} {\bibfnamefont {P.}~\bibnamefont {Audebert}}, \bibinfo {author}
  {\bibfnamefont {E.}~\bibnamefont {Brambrink}}, \bibinfo {author}
  {\bibfnamefont {C.~A.}\ \bibnamefont {Cecchetti}}, \bibinfo {author}
  {\bibfnamefont {A.}~\bibnamefont {Pipahl}}, \bibinfo {author} {\bibfnamefont
  {L.}~\bibnamefont {Romagnani}}, \ and\ \bibinfo {author} {\bibfnamefont
  {O.}~\bibnamefont {Willi}},\ }\href {\doibase 10.1126/science.1124412}
  {\bibfield  {journal} {\bibinfo  {journal} {Science}\ }\textbf {\bibinfo
  {volume} {312}},\ \bibinfo {pages} {410} (\bibinfo {year}
  {2006})}\BibitemShut {NoStop}%
\bibitem [{\citenamefont {Schollmeier}\ \emph {et~al.}(2008)\citenamefont
  {Schollmeier}, \citenamefont {Becker}, \citenamefont {Gei\ss{}el},
  \citenamefont {Flippo}, \citenamefont {Bla\ifmmode \check{z}\else
  \v{z}\fi{}evi\ifmmode~\acute{c}\else \'{c}\fi{}}, \citenamefont {Gaillard},
  \citenamefont {Gautier}, \citenamefont {Gr\"uner}, \citenamefont {Harres},
  \citenamefont {Kimmel}, \citenamefont {N\"urnberg}, \citenamefont {Rambo},
  \citenamefont {Schramm}, \citenamefont {Schreiber}, \citenamefont
  {Sch\"utrumpf}, \citenamefont {Schwarz}, \citenamefont {Tahir}, \citenamefont
  {Atherton}, \citenamefont {Habs}, \citenamefont {Hegelich},\ and\
  \citenamefont {Roth}}]{Schollmeier2008}%
  \BibitemOpen
  \bibfield  {author} {\bibinfo {author} {\bibfnamefont {M.}~\bibnamefont
  {Schollmeier}}, \bibinfo {author} {\bibfnamefont {S.}~\bibnamefont {Becker}},
  \bibinfo {author} {\bibfnamefont {M.}~\bibnamefont {Gei\ss{}el}}, \bibinfo
  {author} {\bibfnamefont {K.~A.}\ \bibnamefont {Flippo}}, \bibinfo {author}
  {\bibfnamefont {A.}~\bibnamefont {Bla\ifmmode \check{z}\else
  \v{z}\fi{}evi\ifmmode~\acute{c}\else \'{c}\fi{}}}, \bibinfo {author}
  {\bibfnamefont {S.~A.}\ \bibnamefont {Gaillard}}, \bibinfo {author}
  {\bibfnamefont {D.~C.}\ \bibnamefont {Gautier}}, \bibinfo {author}
  {\bibfnamefont {F.}~\bibnamefont {Gr\"uner}}, \bibinfo {author}
  {\bibfnamefont {K.}~\bibnamefont {Harres}}, \bibinfo {author} {\bibfnamefont
  {M.}~\bibnamefont {Kimmel}}, \bibinfo {author} {\bibfnamefont
  {F.}~\bibnamefont {N\"urnberg}}, \bibinfo {author} {\bibfnamefont
  {P.}~\bibnamefont {Rambo}}, \bibinfo {author} {\bibfnamefont
  {U.}~\bibnamefont {Schramm}}, \bibinfo {author} {\bibfnamefont
  {J.}~\bibnamefont {Schreiber}}, \bibinfo {author} {\bibfnamefont
  {J.}~\bibnamefont {Sch\"utrumpf}}, \bibinfo {author} {\bibfnamefont
  {J.}~\bibnamefont {Schwarz}}, \bibinfo {author} {\bibfnamefont {N.~A.}\
  \bibnamefont {Tahir}}, \bibinfo {author} {\bibfnamefont {B.}~\bibnamefont
  {Atherton}}, \bibinfo {author} {\bibfnamefont {D.}~\bibnamefont {Habs}},
  \bibinfo {author} {\bibfnamefont {B.~M.}\ \bibnamefont {Hegelich}}, \ and\
  \bibinfo {author} {\bibfnamefont {M.}~\bibnamefont {Roth}},\ }\href {\doibase
  10.1103/PhysRevLett.101.055004} {\bibfield  {journal} {\bibinfo  {journal}
  {Phys. Rev. Lett.}\ }\textbf {\bibinfo {volume} {101}},\ \bibinfo {pages}
  {055004} (\bibinfo {year} {2008})}\BibitemShut {NoStop}%
\bibitem [{\citenamefont {Schillaci}\ \emph
  {et~al.}(2016{\natexlab{a}})\citenamefont {Schillaci}, \citenamefont
  {Maggiore}, \citenamefont {Andó}, \citenamefont {Cirrone}, \citenamefont
  {Cuttone}, \citenamefont {Romano}, \citenamefont {Scuderi}, \citenamefont
  {Allegra}, \citenamefont {Amato}, \citenamefont {Gallo}, \citenamefont
  {Korn}, \citenamefont {Leanza}, \citenamefont {Margarone}, \citenamefont
  {Milluzzo},\ and\ \citenamefont {Petringa}}]{SchillaciJINST}%
  \BibitemOpen
  \bibfield  {author} {\bibinfo {author} {\bibfnamefont {F.}~\bibnamefont
  {Schillaci}}, \bibinfo {author} {\bibfnamefont {M.}~\bibnamefont {Maggiore}},
  \bibinfo {author} {\bibfnamefont {L.}~\bibnamefont {Andó}}, \bibinfo
  {author} {\bibfnamefont {G.}~\bibnamefont {Cirrone}}, \bibinfo {author}
  {\bibfnamefont {G.}~\bibnamefont {Cuttone}}, \bibinfo {author} {\bibfnamefont
  {F.}~\bibnamefont {Romano}}, \bibinfo {author} {\bibfnamefont
  {V.}~\bibnamefont {Scuderi}}, \bibinfo {author} {\bibfnamefont
  {L.}~\bibnamefont {Allegra}}, \bibinfo {author} {\bibfnamefont
  {A.}~\bibnamefont {Amato}}, \bibinfo {author} {\bibfnamefont
  {G.}~\bibnamefont {Gallo}}, \bibinfo {author} {\bibfnamefont
  {G.}~\bibnamefont {Korn}}, \bibinfo {author} {\bibfnamefont {R.}~\bibnamefont
  {Leanza}}, \bibinfo {author} {\bibfnamefont {D.}~\bibnamefont {Margarone}},
  \bibinfo {author} {\bibfnamefont {G.}~\bibnamefont {Milluzzo}}, \ and\
  \bibinfo {author} {\bibfnamefont {G.}~\bibnamefont {Petringa}},\ }\href
  {http://stacks.iop.org/1748-0221/11/i=08/a=P08022} {\bibfield  {journal}
  {\bibinfo  {journal} {Journal of Instrumentation}\ }\textbf {\bibinfo
  {volume} {11}},\ \bibinfo {pages} {P08022} (\bibinfo {year}
  {2016}{\natexlab{a}})}\BibitemShut {NoStop}%
\bibitem [{\citenamefont {Masood}\ \emph {et~al.}(2014)\citenamefont {Masood},
  \citenamefont {Bussmann}, \citenamefont {Cowan}, \citenamefont {Enghardt},
  \citenamefont {Karsch}, \citenamefont {Kroll}, \citenamefont {Schramm},\ and\
  \citenamefont {Pawelke}}]{Masood2014}%
  \BibitemOpen
  \bibfield  {author} {\bibinfo {author} {\bibfnamefont {U.}~\bibnamefont
  {Masood}}, \bibinfo {author} {\bibfnamefont {M.}~\bibnamefont {Bussmann}},
  \bibinfo {author} {\bibfnamefont {T.~E.}\ \bibnamefont {Cowan}}, \bibinfo
  {author} {\bibfnamefont {W.}~\bibnamefont {Enghardt}}, \bibinfo {author}
  {\bibfnamefont {L.}~\bibnamefont {Karsch}}, \bibinfo {author} {\bibfnamefont
  {F.}~\bibnamefont {Kroll}}, \bibinfo {author} {\bibfnamefont
  {U.}~\bibnamefont {Schramm}}, \ and\ \bibinfo {author} {\bibfnamefont
  {J.}~\bibnamefont {Pawelke}},\ }\href {\doibase 10.1007/s00340-014-5796-z}
  {\bibfield  {journal} {\bibinfo  {journal} {Applied Physics B}\ }\textbf
  {\bibinfo {volume} {117}},\ \bibinfo {pages} {41} (\bibinfo {year}
  {2014})}\BibitemShut {NoStop}%
\bibitem [{\citenamefont {Schillaci}\ \emph
  {et~al.}(2016{\natexlab{b}})\citenamefont {Schillaci}, \citenamefont
  {Pommarel}, \citenamefont {Romano}, \citenamefont {Cuttone}, \citenamefont
  {Costa}, \citenamefont {Giove}, \citenamefont {Maggiore}, \citenamefont
  {Russo}, \citenamefont {Scuderi}, \citenamefont {Malka}, \citenamefont
  {Vauzour}, \citenamefont {Flacco},\ and\ \citenamefont
  {Cirrone}}]{SchillaciJINST2}%
  \BibitemOpen
  \bibfield  {author} {\bibinfo {author} {\bibfnamefont {F.}~\bibnamefont
  {Schillaci}}, \bibinfo {author} {\bibfnamefont {L.}~\bibnamefont {Pommarel}},
  \bibinfo {author} {\bibfnamefont {F.}~\bibnamefont {Romano}}, \bibinfo
  {author} {\bibfnamefont {G.}~\bibnamefont {Cuttone}}, \bibinfo {author}
  {\bibfnamefont {M.}~\bibnamefont {Costa}}, \bibinfo {author} {\bibfnamefont
  {D.}~\bibnamefont {Giove}}, \bibinfo {author} {\bibfnamefont
  {M.}~\bibnamefont {Maggiore}}, \bibinfo {author} {\bibfnamefont
  {A.}~\bibnamefont {Russo}}, \bibinfo {author} {\bibfnamefont
  {V.}~\bibnamefont {Scuderi}}, \bibinfo {author} {\bibfnamefont
  {V.}~\bibnamefont {Malka}}, \bibinfo {author} {\bibfnamefont
  {B.}~\bibnamefont {Vauzour}}, \bibinfo {author} {\bibfnamefont
  {A.}~\bibnamefont {Flacco}}, \ and\ \bibinfo {author} {\bibfnamefont
  {G.}~\bibnamefont {Cirrone}},\ }\href@noop {} {\bibfield  {journal} {\bibinfo
   {journal} {Journal of Instrumentation}\ }\textbf {\bibinfo {volume} {11}},\
  \bibinfo {pages} {T07005} (\bibinfo {year} {2016}{\natexlab{b}})}\BibitemShut
  {NoStop}%
\bibitem [{\citenamefont {Fonseca}\ \emph {et~al.}(2002)\citenamefont
  {Fonseca}, \citenamefont {Silva}, \citenamefont {Tsung}, \citenamefont
  {Decyk}, \citenamefont {Lu}, \citenamefont {Ren}, \citenamefont {Mori},
  \citenamefont {Deng}, \citenamefont {Lee}, \citenamefont {Katsouleas},\ and\
  \citenamefont {Adam}}]{Fonseca2002}%
  \BibitemOpen
  \bibfield  {author} {\bibinfo {author} {\bibfnamefont {R.~A.}\ \bibnamefont
  {Fonseca}}, \bibinfo {author} {\bibfnamefont {L.~O.}\ \bibnamefont {Silva}},
  \bibinfo {author} {\bibfnamefont {F.~S.}\ \bibnamefont {Tsung}}, \bibinfo
  {author} {\bibfnamefont {V.~K.}\ \bibnamefont {Decyk}}, \bibinfo {author}
  {\bibfnamefont {W.}~\bibnamefont {Lu}}, \bibinfo {author} {\bibfnamefont
  {C.}~\bibnamefont {Ren}}, \bibinfo {author} {\bibfnamefont {W.~B.}\
  \bibnamefont {Mori}}, \bibinfo {author} {\bibfnamefont {S.}~\bibnamefont
  {Deng}}, \bibinfo {author} {\bibfnamefont {S.}~\bibnamefont {Lee}}, \bibinfo
  {author} {\bibfnamefont {T.}~\bibnamefont {Katsouleas}}, \ and\ \bibinfo
  {author} {\bibfnamefont {J.~C.}\ \bibnamefont {Adam}}\ }(\bibinfo
  {publisher} {Springer Berlin Heidelberg},\ \bibinfo {address} {Berlin,
  Heidelberg},\ \bibinfo {year} {2002})\ pp.\ \bibinfo {pages}
  {342--351}\BibitemShut {NoStop}%
\bibitem [{sim()}]{sim}%
  \BibitemOpen
  \href@noop {} {\bibinfo  {journal} {The simulation resolved this domain with
  $32768 \times 384$ cells, used a time step $\Delta t = 0.058$fs, 16
  particles/species/cell, cubic particle shapes, and ran for $\sim 4$ ps. The
  boundary conditions were open/absorbing in the longitudinal direction and
  periodic transversely}\ }\BibitemShut {NoStop}%
\bibitem [{\citenamefont {Macchi}\ \emph {et~al.}(2012)\citenamefont {Macchi},
  \citenamefont {Nindrayog},\ and\ \citenamefont {Pegoraro}}]{Macchi2012}%
  \BibitemOpen
\bibfield  {journal} {  }\bibfield  {author} {\bibinfo {author} {\bibfnamefont
  {A.}~\bibnamefont {Macchi}}, \bibinfo {author} {\bibfnamefont {A.~S.}\
  \bibnamefont {Nindrayog}}, \ and\ \bibinfo {author} {\bibfnamefont
  {F.}~\bibnamefont {Pegoraro}},\ }\href {\doibase 10.1103/PhysRevE.85.046402}
  {\bibfield  {journal} {\bibinfo  {journal} {Phys. Rev. E}\ }\textbf {\bibinfo
  {volume} {85}},\ \bibinfo {pages} {046402} (\bibinfo {year}
  {2012})}\BibitemShut {NoStop}%
\end{thebibliography}%

\end{document}